\documentclass[12pt]{article}
\usepackage{epsfig,amssymb,amsmath,psfrag}

\newcommand{\cP}{{\cal P}}
\newcommand{\cN}{{\cal N}}

\newcommand{\cO}{{\cal O}}

\def\eqn#1{eq.~(\ref{#1})}
\def\Eqn#1{Equation~(\ref{#1})}
\def\eqns#1#2{eqs.~(\ref{#1}) and~(\ref{#2})}

\def\fig#1{fig.~\ref{#1}}

\def \be  {\begin{equation}}
\def \ee  {\end{equation}}
\def \ba  {\begin{eqnarray}}
\def \ea  {\end{eqnarray}}
\def\bea{\begin{eqnarray}}
\def\eea{\end{eqnarray}}

\def \ln {{\rm ln}}
\def \cO{\mathcal{O}}

\newcommand \Li{{\rm Li}}
\def\polylog{ {\rm Li} }
\def\ln{ \log }

\begin{document}

\thispagestyle{empty}

\begin{center}
CERN--PH--TH/2011/251\hskip0.5cm
SLAC--PUB--14632 \hskip0.5cm
LAPTH-042/11 \hskip0.5cm 
HU-EP-11/44
\end{center}

\begingroup\centering
{\Large\bfseries\mathversion{bold}
Analytic result for the two-loop six-point NMHV amplitude
in $\cN=4$ super Yang-Mills theory
\par}%
\vspace{7mm}

\begingroup\scshape\large
Lance~J.~Dixon$^{(1)}$, James~M.~Drummond$^{(2,3)}$\\
and Johannes M.~Henn$^{(4,5)}$\\
\endgroup
\vspace{8mm}
\begingroup\small
$^{(1)}$ \emph{SLAC National Accelerator Laboratory,
Stanford University, Stanford, CA 94309, USA} \\
$^{(2)}$ \emph{PH-TH Division, CERN, Geneva, Switzerland} \\
$^{(3)}$ \emph{LAPTH, Universit\'e de Savoie, CNRS,
B.P. 110, F-74941 Annecy-le-Vieux Cedex, France}\\
$^{(4)}$ \emph{
Humboldt-Universit\"at
zu Berlin, Newtonstra{\ss}e 15, 12489 Berlin, Germany}\\
$^{(5)}$ \emph{Institute for Advanced Study, 
Princeton, NJ 08540, USA}
\endgroup

\vspace{0.6cm}
\begingroup\small
 {\tt lance@slac.stanford.edu}\,,\; {\tt drummond@lapp.in2p3.fr}\,,\; {\tt jmhenn@ias.edu}
\endgroup
\vspace{1.2cm}

\textbf{Abstract}\vspace{5mm}\par
\begin{minipage}{14.7cm}
We provide a simple analytic formula for the two-loop six-point ratio
function of planar $\cN=4$ super Yang-Mills theory. This result
extends the analytic knowledge of multi-loop six-point amplitudes
beyond those with maximal helicity violation.  We make a natural
ansatz for the symbols of the relevant functions appearing in the
two-loop amplitude, and impose various consistency conditions,
including symmetry, the absence of spurious poles, the correct
collinear behaviour, and agreement with the operator product expansion
for light-like (super) Wilson loops.  This information reduces the
ansatz to a small number of relatively simple functions.  In order to
fix these parameters uniquely, we utilize an explicit representation
of the amplitude in terms of loop integrals that can be evaluated
analytically in various kinematic limits.  The final compact analytic
result is expressed in terms of classical polylogarithms, whose
arguments are rational functions of the dual conformal cross-ratios,
plus precisely two functions that are not of this type.  One of the
functions, the loop integral $\Omega^{(2)}$, also plays a key role in
a new representation of the remainder function $\mathcal{R}_6^{(2)}$
in the maximally helicity violating sector.  Another interesting
feature at two loops is the appearance of a new (parity odd) $\times$
(parity odd) sector of the amplitude, which is absent at one loop, and
which is uniquely determined in a natural way in terms of the more
familiar (parity even) $\times$ (parity even) part.  The second
non-polylogarithmic function, the loop integral
$\tilde{\Omega}^{(2)}$, characterizes this sector.  Both
$\Omega^{(2)}$ and $\tilde{\Omega}^{(2)}$ can be expressed as
one-dimensional integrals over classical polylogarithms with rational
arguments.
\end{minipage}\par
\endgroup

\newpage

\tableofcontents

\section{Introduction}
\setcounter {equation} {0}

Much progress has been achieved recently in the analytic understanding of
seemingly complicated scattering processes. In particular, attention has
been focused on the planar sector, or large $N$ limit, of maximally
supersymmetric $\cN=4$ Yang-Mills theory.
The scattering amplitudes in this sector of the theory obey many
startling properties, which has led to the hope that the general scattering
problem might be solvable, exactly in the coupling.

One of the major simplifications that the planar $\cN=4$ theory
enjoys is dual conformal
symmetry~\cite{Drummond:2006rz,Bern:2006ew,Drummond:2007aua,%
Drummond:2007cf,Drummond:2007au,Alday:2007hr,Alday:2007he}, 
which dictates how colour-ordered
amplitudes behave under conformal transformations of the dual (or region)
variables defined via $p_i = x_i - x_{i+1}$. For the particular case of
maximally helicity-violating (MHV) amplitudes, this symmetry is intimately
connected to the relation between the amplitudes and Wilson
loops evaluated on polygons with light-like edges, whose vertices are
located at the
$x_i$~\cite{Alday:2007hr,Drummond:2007aua,Brandhuber:2007yx,%
Drummond:2007cf,Drummond:2007au,Drummond:2007bm,Bern:2008ap,%
Drummond:2008aq}.
The MHV amplitudes are infrared divergent, just as the
corresponding light-like Wilson loops are ultraviolet divergent.
The Wilson-loop divergence has the consequence that a suitably-defined
finite part transforms anomalously under the dual conformal
symmetry~\cite{Drummond:2007cf,Drummond:2007au}.
The Ward identity describing this behaviour actually fixes the form of
the four-point and five-point amplitudes to all orders in the coupling,
to that given by the BDS ansatz~\cite{Bern:2005iz}.
From six points onwards, the existence of dual-conformal invariant
cross-ratios means that the problem of determining the MHV amplitude
reduces to finding a function that depends only on the cross-ratios ---
the so-called `remainder function', which corrects the BDS ansatz.

Great advances have
been made recently in understanding the form of the remainder function,
which is non-trivial beginning at two loops.
The need for a two-loop remainder function
for Wilson loops was observed for a large number of points
in ref.~\cite{Alday:2007he}, and for six points in
ref.~\cite{Drummond:2007bm}.  The multi-Regge limit of the six-point
scattering amplitude also implied a non-trivial remainder
function~\cite{Bartels:2008ce}.  At a few generic kinematic points, the
Wilson loop~\cite{Drummond:2008aq} and amplitude~\cite{Bern:2008ap}
remainder functions were found to agree numerically.
The six-point Wilson loop integrals entering the remainder function
were computed analytically in terms of Goncharov
polylogarithms~\cite{DelDuca:2009au,DelDuca:2010zg},
and then simplified down to classical
polylogarithms~\cite{Goncharov:2010jf}
using the notion of the {\it symbol} of a pure
function~\cite{symbolsC,symbolsB,symbolsG,Goncharov:2010jf}.
The integrals contributing to the six-point MHV scattering
amplitude have also been evaluated analytically~\cite{Drummond:2010mb}
in a certain kinematical regime using a mass regulator~\cite{Alday:2009zm},
and the remainder function has been found to agree
with the Wilson-loop expression of ref.~\cite{Goncharov:2010jf}.
Very recently, the symbol for the three-loop six-point remainder function
was determined up to two arbitrary parameters~\cite{Dixon:2011pw},
by imposing a variety of constraints, in particular the operator product
expansion (OPE) for Wilson loops developed in
refs.~\cite{Alday:2010ku,Gaiotto:2010fk,Gaiotto:2011dt}.

For more than six points, numerical results for the remainder function
have been obtained via Wilson loop
integrals~\cite{Anastasiou:2009kna,Brandhuber:2009da}.
Integral representations for the MHV amplitudes have been
presented at seven points~\cite{Vergu:2009zm} and for an arbitrary number
of points using momentum twistors~\cite{ArkaniHamed:2010gh}.
Recently, an expression for the symbol of the two-loop remainder function 
has been given for an arbitrary number of points~\cite{CaronHuot:2011ky},
and the structure of the OPE for this case has been
explored~\cite{Sever:2011pc}.
In special kinematics corresponding to scattering in two space-time
dimensions, analytic results are available for a number of configurations
at two loops~\cite{DelDuca:2010zp,Heslop:2010kq,Alday:2010jz}
and conjecturally even at three loops~\cite{Heslop:2011hv}.

When one considers amplitudes beyond the MHV sector, there is another
finite dual conformally invariant quantity that one can consider, namely
the `ratio function' $\cP$. This quantity is defined by factoring
out the MHV superamplitude from the full
superamplitude~\cite{Drummond:2008vq}, 
\be
\mathcal{A} = \mathcal{A}_{\rm MHV} \, \times \, \cP\,.
\label{ratiofunctiondef}
\ee
Infrared divergences are universal for all component amplitudes;
hence the MHV factor contains all such divergences, leaving an
infrared finite quantity $\cP$. One of the central conjectures of 
ref.~\cite{Drummond:2008vq} is that $\cP$ is also dual conformally
invariant.  There is strong supporting evidence for this conjecture in
the form of direct analytic one-loop
results~\cite{Bern:1994cg,Bern:2004ky,Huang:2005ve,Risager:2005ke,%
Drummond:2008bq} and, in the six-point case, numerical evidence at two
loops~\cite{Kosower:2010yk}.  In this paper, we will
construct the ratio function $\cP$ analytically at two loops for
six external legs.

At tree level, the ratio function is given by a sum over dual
superconformal `$R$-invariants' \cite{Drummond:2008vq,Drummond:2008cr,%
ArkaniHamed:2009dn,Mason:2009qx,ArkaniHamed:2009vw,Korchemsky:2010ut}.
These quantities are invariant under a much larger
(infinite-dimensional) Yangian symmetry, obtained by combining
invariance under both the original and dual copies of superconformal
symmetry~\cite{Drummond:2009fd}. Beyond tree level one finds
$R$-invariants dressed by dual conformally invariant
functions~\cite{Drummond:2008vq,Drummond:2008bq,Kosower:2010yk}.
Since the $R$-invariants individually exhibit spurious poles, which
cannot appear in the final amplitude, they cannot appear in an
arbitrary way.  The particular linear combination appearing in the
tree amplitude is free of spurious poles.  At loop level, the absence
of spurious poles implies restrictions on the dual conformally
invariant functions that dress them~\cite{Korchemsky:2009hm}.
Additional restrictions on these same functions come from the known
behaviour of the amplitude when two of the external particles become
collinear. These constraints will be important in our construction of
$\cP$ at two loops.

A consequence of the duality between MHV amplitudes and light-like Wilson
loops is that the remainder function can be analysed by conformal field
theory methods, such as the operator product expansion
(OPE)~\cite{Alday:2010ku,Gaiotto:2010fk,Gaiotto:2011dt}. 
Various proposals have been put forward for extending the duality between
amplitudes and Wilson loops beyond the MHV sector, either in terms of a
supersymmetric version of the Wilson
loop~\cite{Mason:2010yk,CaronHuot:2010ek,CaronHuot:2011ky}, or in terms of
correlation functions~\cite{Eden:2011yp,Eden:2011ku}. Although there may
be various subtleties in realising such an object, compatible with the
full $\cN=4$ supersymmetry in a Lagrangian
formulation~\cite{Belitsky:2011zm}, one may instead justify
the existence of such an object through the OPE. The framework for
pursuing this approach was developed in ref.~\cite{Sever:2011da},
and agreement was found with the known one-loop six-point next-to-MHV (NMHV)
amplitude~\cite{Bern:1994cg,Huang:2005ve,Risager:2005ke,Drummond:2008bq}.
This agreement provides non-trivial evidence that there
does indeed exist a Wilson-loop quantity dual to all scattering amplitudes.

The aim of this paper is to combine various approaches in order to
determine the six-point ratio function $\cP$, or equivalently the NMHV
amplitude, analytically at two loops.  This quantity was expressed
in terms of dual conformal integrals, and computed numerically,
in ref.~\cite{Kosower:2010yk}.  We proceed in a manner similar
to our recent examination of the three-loop six-point remainder
function~\cite{Dixon:2011pw}.  In particular, we make an ansatz for the
symbols of the various pure functions involved. (See
appendix~\ref{app-symbols} for a brief introduction to pure functions and
their symbols.)  In other words, we assume that the functions that appear
fall within a particular class of multi-dimensional iterated integrals, or
generalized polylogarithms.  We say functions rather than function because
in general, beyond one loop, one can imagine that there are both (parity
even)$\times$(parity even) and (parity odd)$\times$(parity odd)
contributions, in a sense which we make specific in the next section.  For
convenience, we call these contributions `even' and `odd',
respectively.  At tree level and at one loop, the odd part vanishes.

After constructing an ansatz, the next step is to impose consistency
conditions.  We first impose the spurious pole and collinear conditions.
Then we impose that a certain double discontinuity is compatible with the
OPE~\cite{Alday:2010ku,Gaiotto:2010fk,Gaiotto:2011dt,Sever:2011da}.  At
this stage, we find that the symbols for the relevant functions contain
nine unfixed parameters.  We convert these symbols into explicit
functions.  In general, this step leads to `beyond-the-symbol
ambiguities'.  These ambiguities are associated with functions whose
symbols vanish identically, namely transcendental constants, such as the
Riemann $\zeta$ values $\zeta_p$, multiplied by pure functions of lower
degree. However, in the present case, after re-imposing the spurious
and collinear restrictions at the level of functions, there is only
one additional ambiguity, associated with adding the product of $\zeta_2$
with the one-loop ratio function.  This term obeys all constraints by
itself and has vanishing symbol. We are thus left with a ten-dimensional
space of functions. In particular, we find that the odd part is
necessarily non-zero.  Moreover, it is uniquely determined in terms of the
even part.

In order to fix the remaining free parameters, we turn to 
a representation of the even part of the
two-loop six-point NMHV amplitude based on loop
integrals~\cite{Kosower:2010yk}.  We analyse this representation,
appropriately rewritten with a mass regulator~\cite{Alday:2009zm}, in the
symmetric regime with all three cross-ratios equal to $u$.  In this
regime, the most cumbersome double-pentagon integrals can be traded for
the MHV remainder function, plus simpler integrals.  This observation
allows us to perform an analytic expansion for small and large $u$.
Comparing these expansions with the ansatz, we are able to match them,
precisely fixing all remaining free parameters.  The fact that the ansatz
agrees with the expansion of the loop-integral calculation in this regime
is a highly non-trivial cross check, since an entire function is matched
by an ansatz with just a few free parameters.  Further confirmation that our
result is correct comes from comparing with a numerical
evaluation~\cite{Kosower:2010yk} at a particular asymmetric
kinematical point. This latter check also confirms the expectation
that $\cP$ is defined independently of any infrared
regularization scheme. See also ref.~\cite{Henn:2011by} for a recent
discussion of different infrared regularizations and regularization-scheme
independence.

In contrast to the the two-loop six-point MHV
amplitude~\cite{Goncharov:2010jf}, the two-loop six-point ratio
function cannot quite be expressed in terms of classical
polylogarithms.  Two additional functions appear,
one in the even part and one in the odd part.
However, these functions have a very simple structure:
we can write them as simple one-dimensional integrals over classical
polylogarithmic functions of degree three.  The even part of the ratio
function can be written in terms of single-variable polylogarithmic
functions whose arguments are rational in the three cross-ratios $u,v,w$,
plus one of the new functions, which coincides with the finite
double-pentagon integral $\Omega^{(2)}$~\cite{ArkaniHamed:2010kv}.
We use the differential equations obeyed by this
integral~\cite{Drummond:2010cz,Dixon:2011ng} to derive various parametric
integral representations for it. The odd part consists entirely
of the second new function, $\tilde{V}$, which also can be expressed as
a single integral over classical polylogarithms of degree three.
This function can also be identified as the odd part of another finite
double-pentagon integral,
$\tilde{\Omega}^{(2)}$~\cite{ArkaniHamed:2010kv}, which we compute using 
the differential equations derived in ref.~\cite{Drummond:2010cz}.

This paper is organized as follows. In section~\ref{def-nmhv} we review
non-MHV amplitudes, and the definition of the ratio function in planar
$\cN=4$ super Yang-Mills theory.  We discuss the physical constraints
satisfied by $\cP$, namely in the collinear and spurious limits, and also
those arising from the OPE expansion of super Wilson loops.  We make an ansatz
for the symbol of $\cP$ at two loops in section~\ref{sect-ansatz}, and
then apply the constraints.  
In order to promote the symbol to a function, 
we introduce in section~\ref{sect-symboltofunction1} two new
functions that are not expressible in terms of
classical polylogarithms, but have simple parametric integral
definitions. 
Next, in section~\ref{sect-symboltofunction2},
we parametrise the beyond-the-symbol ambiguities
and apply the collinear and spurious constraints at the
functional level, which leaves only ten unfixed parameters. In
section~\ref{sect-analytic} we determine these parameters by performing
an analytic two-loop evaluation of the integrals contributing to the
even part of the NMHV amplitude in a special kinematical regime.  
The final result for the full two-loop NMHV ratio
function is presented in section \ref{sect-final}. We conclude in section
\ref{sect-conclusions}.  Several appendices contain background material
and technical details.   We provide the symbols for several
of the quantities appearing in this article as auxiliary material.


\section{Non-MHV amplitudes and the ratio function}
\label{def-nmhv}
\setcounter {equation} {0}

To describe the scattering amplitudes of $\cN=4$ super Yang-Mills theory,
it is useful to introduce an on-shell superspace (see {\it e.g.}
refs.~\cite{Nair:1988bq,Georgiou:2004by,Drummond:2008vq,ArkaniHamed:2008gz}).
All the different on-shell states of the theory can be arranged into an
on-shell superfield $\Phi$ which depends on Grassmann variables $\eta^A$
transforming in the fundamental representation of $su(4)$,
\be
\Phi = G^+ + \eta^A \Gamma_A + \tfrac{1}{2!} \eta^A \eta^B S_{AB}
+ \tfrac{1}{3!} \eta^A \eta^B \eta^C \epsilon_{ABCD} \overline{\Gamma}^D
+ \tfrac{1}{4!} \eta^A \eta^B \eta^C \eta^D \epsilon_{ABCD} G^-.
\label{onshellmultiplet}
\ee
Here $G^+$, $\Gamma_A$,
$S_{AB}=\tfrac{1}{2}\epsilon_{ABCD}\overline{S}^{CD}$,
$\overline{\Gamma}^A$, and $G^-$ are the positive-helicity gluon,
gluino, scalar, anti-gluino, and negative-helicity gluon states,
respectively.  These on-shell states carry a definite null momentum,
which can be written in terms of two commuting spinors,
$p^{\alpha \dot\alpha} =\lambda^\alpha \tilde{\lambda}^{\dot\alpha}$.
Note that the spinors $\lambda$ and $\tilde{\lambda}$ are not uniquely
defined, given $p$; they can be rescaled by $\lambda \rightarrow c\lambda$,
$\tilde{\lambda} \rightarrow c^{-1} \tilde{\lambda}$.  The transformation
properties of the states and the $\eta$ variables are such that the full
superfield has weight 1 under the following operator,
\be
h = -\frac{1}{2}\Bigl[
  \lambda^\alpha\frac{\partial}{\partial \lambda^\alpha}
 - \tilde{\lambda}^{\dot\alpha}
  \frac{\partial}{\partial \tilde{\lambda}^{\dot\alpha}}
 - \eta^A \frac{\partial}{\partial \eta^A}\Bigr]\,.
\ee

All the different (colour-ordered) scattering amplitudes of the theory are
then combined into a single superamplitude
$\mathcal{A}(\Phi_1,\Phi_2,\ldots,\Phi_n)$, from which individual
components can be extracted by expanding in the Grassmann variables
$\eta_i^A$ associated to the different particles. The tree-level MHV
superamplitude is the simplest cyclically invariant quantity with the
correct scaling behaviour for each particle that manifests translation
invariance and supersymmetry,
\be \label{superMHVtree}
\mathcal{A}_{\rm MHV}^{(0)} = i \,
\frac{\delta^4(p) \delta^8(q)}{\langle 1 2 \rangle \langle 2 3 \rangle
                        \cdots \langle n 1 \rangle} \,.
\ee
The arguments of the delta functions are the total momentum
$p^{\alpha \dot\alpha}
= \sum_i \lambda_i^\alpha \tilde{\lambda}_i^{\dot\alpha}$ and
total chiral supercharge $q^{\alpha A} = \sum_i \lambda_i^\alpha \eta_i^A$,
respectively. The full MHV superamplitude is the tree-level one multiplied
by an infrared-divergent factor,
\be
\mathcal{A}_{\rm MHV} = \mathcal{A}_{\rm MHV}^{(0)} \, \times \, M \,.
\label{fullMHV}
\ee

Moving beyond MHV amplitudes, we define the ratio function by factoring
out the MHV superamplitude from the full
superamplitude~\cite{Drummond:2008vq},
\be
\mathcal{A} = \mathcal{A}_{\rm MHV} \, \times \, \cP\,.
\label{ratiodef}
\ee
Here $\cP$ has an expansion in terms of increasing Grassmann
degree, corresponding to the type of amplitudes (MHV, NMHV, N${}^2$MHV,
{\it etc.}),
\be
\cP = 1 + \cP_{\rm NMHV}
+ \cP_{\rm N{}^2MHV} + \ldots
+ \cP_{\overline{\rm MHV}}\,.
\ee
The number of terms in the above expansion of $\cP$ is $(n-3)$,
where $n$ is the number of external legs.
The Grassmann degrees of the terms are $0,4,8,\ldots,(4n-16)$.
At six points, which is the case of interest for this paper,
there are just three terms, corresponding to MHV, NMHV and N${}^2$MHV.
The N${}^2$MHV amplitudes for $n=6$ are equivalent to
$\overline{\rm MHV}$ amplitudes, which are simply related to the MHV
amplitudes by parity.  Thus the non-trivial content
of the ratio function at six points is in the NMHV term.

At tree level, $\cP$ is given by a sum over dual
superconformal `$R$-invariants'~\cite{Drummond:2008vq}.  In particular,
for six points we have
\be
\cP^{(0)}_{\rm NMHV} = R_{1;35} + R_{1;36} + R_{1;46}\,.
\label{Ptree}
\ee
The $R$-invariants can be described using dual coordinates
$x_i$, $\theta_i$ defined by
\be
p_i^{\alpha \dot\alpha} = 
\lambda_i^\alpha \tilde{\lambda}_i^{\dot\alpha}
= x_i^{\alpha \dot\alpha} - x_{i+1}^{\alpha \dot\alpha}, \qquad
q_i^{\alpha A} = \lambda_i^{\alpha} \eta_i^A
= \theta_i^{\alpha A} - \theta_{i+1}^{\alpha A}\,.
\label{xthetadef}
\ee
Then we have~\cite{Drummond:2008vq,Drummond:2008cr}
\be
R_{r;ab} = \frac{\langle a, \, a-1 \rangle \langle b, \, b-1 \rangle
\ \delta^4\bigl( \langle r|x_{ra}x_{ab} |\theta_{br}\rangle
                + \langle r|x_{rb}x_{ba} |\theta_{ar}\rangle \bigr)}
{ x_{ab}^2 \, \langle r|x_{ra} x_{ab} |b\rangle \, 
     \langle r|x_{ra} x_{ab} |b-1 \rangle \, 
     \langle r|x_{rb} x_{ba} |a\rangle \,
     \langle r|x_{rb} x_{ba} |a-1 \rangle } \,.
\label{Rinvdualcoords}
\ee

The $R$-invariants take an even simpler form in terms of momentum
twistors~\cite{Hodges:2009hk,Mason:2009qx}.  These variables are 
(super)twistors associated to the dual space with coordinates
$x,\theta$.  They are defined by
\be
\mathcal{Z}_i = (Z_i \, | \, \chi_i), \qquad 
Z_i^{R=\alpha,\dot\alpha} = 
(\lambda_i^\alpha , x_i^{\beta \dot\alpha}\lambda_{i\beta}),
\qquad
\chi_i^A= \theta_i^{\alpha A}\lambda_{i \alpha} \,.
\ee
The momentum (super)twistors $\mathcal{Z}_i$ transform linearly under
dual (super) conformal symmetry, so that
$(abcd) = \epsilon_{RSTU} Z_a^R Z_b^S Z_c^T Z_d^U$
is a dual conformal invariant. The $R$-invariants can then be written in
terms of the following structures:
\be
[abcde] = 
\frac{\delta^4\bigl(\chi_a (bcde) + {\rm cyclic}\bigr)}
{(abcd)(bcde)(cdea)(deab)(eabc)}\,,
\label{five_bracket_def}
\ee
which contain five terms in the sum over cyclic permutations
of $a,b,c,d,e$ in the delta function.
The bracket notation serves to make clear the totally anti-symmetrised
dependence on five momentum supertwistors. The quantity $R_{r;ab}$ is a
special case of this general invariant,
\be
R_{r;ab} = [r, \, a-1, \, a, \, b-1, \, b]\,.
\label{R5bracket}
\ee
At the six-point level it is clear that there are six different such
invariants.  We label them compactly by $(t)$, using the momentum twistor
$t$ that is absent from the five arguments in the brackets:
\be
(1) \equiv [23456],
\label{compact5bracket}
\ee
and so on. 

In general $R$-invariants obey many identities; see for example
refs.~\cite{Drummond:2008vq,Drummond:2008bq}. These identities can be
organised as residue theorems in the Grassmannian
interpretation~\cite{ArkaniHamed:2009dn}.  At six points, the only
identity we need is~\cite{Drummond:2008vq}
\be
(1)-(2)+(3)-(4)+(5)-(6)=0.
\label{5bracketidentity}
\ee
Using eqs.~(\ref{R5bracket}), (\ref{compact5bracket}) and
(\ref{5bracketidentity}), we can rewrite the NMHV tree
amplitude~(\ref{Ptree}) as
\be
\cP^{(0)}_{\rm NMHV} = [12345] + [12356] + [13456]
 = (6) + (4) + (2) = (1) + (3) + (5).
\label{NewPtree}
\ee

Beyond tree level, the $R$-invariants in the ratio function are dressed
by non-trivial functions of the dual conformal
invariants~\cite{Drummond:2008vq}.  In the six-point case, there are
three independent invariants. We may parametrise the invariants by the
cross-ratios,
\be
u = \frac{x_{13}^2 x_{46}^2}{x_{14}^2 x_{36}^2} \,, \qquad
v = \frac{x_{24}^2 x_{51}^2}{x_{25}^2 x_{41}^2} \,, \qquad
w = \frac{x_{35}^2 x_{62}^2}{x_{36}^2 x_{25}^2} \,.
\label{uvwdef}
\ee
Often it will also be useful to use the variables $y_u,y_v,y_w$ defined by,
\be
y_u = \frac{u-z_+}{u-z_-}, \qquad
y_v = \frac{v-z_+}{v-z_-}, \qquad
y_w = \frac{w-z_+}{w-z_-},
\label{ydef0}
\ee
where
\be
z_\pm = \frac{1}{2}\Bigl[ -1+u+v+w \pm \sqrt{\Delta} \Bigr], \qquad
\Delta = (1-u-v-w)^2 - 4 uvw\,.
\label{zpmDeltadef}
\ee
In terms of momentum twistors, the cross-ratios are expressed as
\be
u = \frac{(6123)(3456)}{(6134)(2356)}, \qquad
v= \frac{(1234)(4561)}{(1245)(3461)}, \qquad
w = \frac{(2345)(5612)}{(2356)(4512)}\,,
\ee
while the $y$ variables simplify to
\be
y_u = \frac{(1345)(2456)(1236)}{(1235)(3456)(1246)}, \quad
y_v = \frac{(1235)(2346)(1456)}{(1234)(2456)(1356)},\quad
y_w = \frac{(2345)(1356)(1246)}{(1345)(2346)(1256)}\,.
\label{ydef1}
\ee
In this form, it is clear that a cyclic rotation by one unit
$Z_i \longrightarrow Z_{i+1}$ maps the $y$ variables as follows,
\be
y_u \longrightarrow \frac{1}{y_v}, \qquad
y_v \longrightarrow \frac{1}{y_w}, \qquad
y_w \longrightarrow \frac{1}{y_u}\,,
\ee
while the cross-ratios behave in the following way,
\be
u \longrightarrow v, \qquad v \longrightarrow w, \qquad w \longrightarrow u\,.
\ee
The parity operation which swaps the sign of the square root of $\Delta$
({\it i.e.}~inverts the $y$ variables) is equivalent to a rotation by three
units in momentum twistor language. Indeed one can think of the
cross-ratios as independent, parity-invariant combinations of the $y$
variables. Specifically we have
\be
u = \frac{y_u (1 - y_v) (1 - y_w)}{(1 - y_u y_v) (1 - y_u y_w)}\,,
\qquad
1-u = \frac{(1-y_u) (1-y_u y_v y_w)}{(1 - y_u y_v) (1 - y_u y_w)} \,,
\label{u1mufromy}
\ee
and similar relations obtained by cyclic rotation.
Because of the ambiguity associated with the sign of the square root of
$\Delta$ in \eqn{ydef0}, the primary definition of the $y$ variables is
through the momentum twistors and \eqn{ydef1}.  Further relations
between these variables are provided in appendix~\ref{sect-variables}.

At six points it can also be convenient to simplify the
momentum-twistor four-brackets by introducing~\cite{Goncharov:2010jf}
antisymmetric two-brackets of $\mathbb{CP}^1$ variables $w_i$
via~\cite{Eden:2011ku}
\be \label{deftwobrackets}
(ij) = \tfrac{1}{4!}\epsilon_{ijklmn}(klmn)\,,
\ee
so that we have
\be
u = \frac{(12)(45)}{(14)(25)}, \qquad
1-u = \frac{(24)(15)}{(14)(25)}, \qquad
y_u = \frac{(26)(13)(45)}{(46)(12)(35)}\,,
\ee
plus six more relations obtained by cyclic permutations.\footnote{%
In comparison with ref.~\cite{Dixon:2011pw}, the indexing of the $w_i$
variables differs by one unit, and a square-root ambiguity in defining the
$y$ variables was resolved in the opposite way.}

Having specified our notation for the invariants we need, we now
parametrise the six-point NMHV ratio function in the following way,
\begin{align}
\cP_{\rm NMHV} = \frac{1}{2}\Bigl[\phantom{a} 
 &[ (1) + (4) ] V_3 + [ (2) + (5) ] V_1 + [ (3) + (6) ] V_2  \notag \\
+\ &[ (1) - (4) ] \tilde{V}_3 - [ (2) - (5) ] \tilde{V}_1
           + [ (3) - (6) ] \tilde{V}_2 \Bigr] \,.
\label{ratio123}
\end{align}
The $V_i$ and $\tilde{V}_i$ are functions of the conformal invariants
and of the coupling, with the $V_i$ even under parity while the
$\tilde{V}_i$ are odd (recall that parity is equivalent to a rotation
by three units). The cyclic and reflection symmetries of the amplitude
$\mathcal{A}$ (and hence the ratio function $\cP$) mean that the $V_i$
and $\tilde{V}_i$ are not all independent. Indeed, choosing
$V_3 = V(u,v,w)$ and $\tilde{V}_3=\tilde{V}(y_u,y_v,y_w)$, we can write
\begin{align}
\cP_{\rm NMHV} =\frac{1}{2}\Bigl[ \phantom{a} 
 &[(1) + (4)] V(u,v,w) + [(2) + (5)] V(v,w,u) + [(3) + (6)] V(w,u,v) \notag \\
+\ &[(1) - (4)] \tilde{V}(y_u,y_v,y_w) - [(2)-(5)] \tilde{V}(y_v,y_w,y_u)
  + [(3) - (6)] \tilde{V}(y_w,y_u,y_v) \Bigr]\,.
\label{PVform}
\end{align}

The functions $V$ and $\tilde{V}$ obey the symmetry properties,
\be
V(w,v,u) = V(u,v,w)\,, \qquad
\tilde{V}(y_w,y_v,y_u) =  - \tilde{V}(y_u,y_v,y_w)\,.
\label{sym}
\ee
Note that we have written the parity-odd function $\tilde{V}$ as a
function of the $y$ variables, while $V$, being parity even, can be
written as a function of the cross-ratios.  The functions $V$ and
$\tilde{V}$ depend on the coupling. We expand them perturbatively
as follows,
\be
V(a) = \sum_{l=0}^\infty a^l V^{(l)}, \qquad
\tilde{V}(a) = \sum_{l=0}^\infty a^l \tilde{V}^{(l)}\,.
\ee
Here
\be
a \equiv \frac{g^2 N}{8 \pi^2} \,,
\label{a_def}
\ee
where $g$ is the Yang-Mills coupling constant for gauge group $SU(N)$;
the planar limit is $N\to\infty$ with $a$ held fixed.

At tree level, we have $V^{(0)}=1$ while $\tilde{V}^{(0)}$
vanishes. One can see that the expression~(\ref{PVform}) with $V=1$
agrees with \eqn{NewPtree}.
At one loop, $\tilde{V}$ still vanishes, while $V$ is a non-trivial
function involving logarithms and dilogarithms,
\bea
V^{(1)} &=& \frac{1}{2} \Bigl[ -\log u \log w + \log(uw) \log v
 + {\rm Li}_2(1-u) + {\rm Li_2}(1-v) + {\rm Li}_2(1-w) - 2 \zeta_{2}
 \Bigr]\,, \nonumber\\
&~& \label{Voneloop}\\
\tilde{V}^{(1)} &=& 0 \,. \label{tildeVoneloop}
\eea
The main results of this paper are analytical two-loop expressions
for $V^{(2)}$ and $\tilde{V}^{(2)}$, both of which are non-vanishing.

Let us discuss some general constraints that the functions $V$ and
$\tilde{V}$ obey.

Physical poles in amplitudes are associated with singular factors
in the denominator involving sums of color-adjacent momenta, of the form
$(p_i + p_{i+1} + \ldots + p_{j-1})^2 \equiv x_{ij}^2$.
In the $R$-invariants, in the notation of \eqn{five_bracket_def},
such poles appear as four brackets $(abcd)$ of the form
\be
(i-1,i,j-1,j) = \langle i-1,i\rangle \, \langle j-1,j\rangle \, x_{ij}^2 \,.
\label{fourbracketphysical}
\ee
However, the $R$-invariants also contain spurious poles, which arise from
the four brackets $(abcd)$ that are not of
this form.  The full amplitude must not have such poles.  Therefore
the functions $V$ and $\tilde{V}$ must conspire to cancel
the pole with a zero in the corresponding kinematical configuration. 

In the dual-coordinate notation~(\ref{Rinvdualcoords}), the $R$-invariants
contain poles from denominator factors of the form
$\langle r |x_{ra} x_{ab}|b\rangle$.
For special values of $a,b,r$, such factors can simplify into physical
singularities, but for generic values they correspond to spurious poles. In
the six-point case, for example, $R_{1;46}$ contains a factor of
\be
   \langle 1 | x_{14}x_{46} | 5 \rangle
 = \langle 1 | x_{14} |4] \,  \langle 4 5 \rangle
\ee
in the denominator. While the pole at $\langle 45 \rangle=0$ is a 
physical (collinear) singularity, the pole at $\langle 1 | x_{14} |4] = 0$
is spurious. In momentum-twistor notation, the spurious pole comes from any
four-bracket in the denominator which is not of the form $(i-1,i,j-1,j)$.

For example, in the six-point case the $R$-invariants $(1)$ and
$(3)$ both contain the spurious factor $(2456)$ in the denominator. 
(In the dual-coordinate notation, this particular pole is proportional to
$\langle 2 | x_{25} |5]$ rather than $\langle 1 | x_{14} |4]$.)
In the tree-level amplitude~(\ref{NewPtree}) there is a cancellation
between the two terms, so we see that 
\be 
(1) \approx - (3)\ \hbox{as}\ (2456)\to 0.
\ee
At loop level, using this relation, we find that the absence of the 
spurious pole implies the following condition on $V$ and $\tilde{V}$,
\be
[V(u,v,w) - V(w,u,v)
 + \tilde{V}(y_u,y_v,y_w) - \tilde{V}(y_w,y_u,y_v)]_{(2456)=0} = 0\,.
\label{spurious}
\ee
As the spurious bracket $(2456)$ vanishes, we find the following
limiting behaviour,
\be
w\rightarrow1\,, \qquad 
y_u \rightarrow (1-w) \frac{u(1-v)}{(u-v)^2}\,,\qquad 
y_v \rightarrow \frac{1}{(1-w)} \frac{(u-v)^2}{v(1-u)}\,, \qquad
y_w \rightarrow \frac{1-u}{1-v}\,.
\ee
This is easiest to see in the two-bracket notation of \eqn{deftwobrackets}, 
in which $(2456)=0$ corresponds to $(13)=0$ and hence to $w_1=w_3$.
The above condition reduces to the one of
ref.~\cite{Korchemsky:2009hm} on the assumption that
$\tilde{V}=0$.\footnote{This is true after correcting a typo in
eq. (3.63) of that reference.}
The one-loop expression for $V$, \eqn{Voneloop}, satisfies the above
constraint with $\tilde{V}^{(1)} = 0$, since $\log w \to 0$ in the limit.

There is also a constraint from the collinear behaviour.  There are two
types of collinear limits, a `$k$-preserving' one where N${}^k$MHV
superamplitudes are related to N${}^k$MHV superamplitudes with one fewer
leg, and a `$k$-decreasing' one which relates N${}^k$MHV superamplitudes
to N${}^{k-1}$MHV superamplitudes with one fewer leg. These two operations
are related to each other by parity and correspond to a
supersymmetrisation of the two splitting functions found when analysing
pure gluon amplitudes~\cite{Mangano:1990by,Bern:1994zx,Bern:1994cg,%
Kosower:1999rx,Kosower:2010yk}.
For the six-point NMHV case, we only need to examine one of the collinear
limits; the other will follow automatically by parity.

Under the collinear limit, the $n$-point amplitude should reduce to the
$(n-1)$-point one multiplied by certain splitting functions. The splitting
functions are automatically taken care of by the MHV prefactor in
\eqn{ratiodef}. The $n$-point ratio function $\cP$ should then be smoothly
related to the $(n-1)$-point one.  Consequently, in the
collinear limit the loop corrections to the six-point ratio function
should vanish, because the five-point ratio function (containing
only MHV and $\overline{\rm MHV}$ components) is exactly equal to its
tree-level value. The $R$-invariants behave smoothly in the limit, either
vanishing or reducing to lower-point invariants. In the case at hand we
can consider the limit $\mathcal{Z}_6 \rightarrow \mathcal{Z}_1$,
which also corresponds to $w_6 \to w_1$, or $x_{35}^2 \to 0$,
or $w\to0$ with $v \to 1-u$.  In this limit,
all $R$-invariants vanish except for $(6)$ and $(1)$, which become
equal. Beyond tree level, the sum of their coefficients must therefore
vanish in the collinear regime. This implies the constraint,
\be
[V(u,v,w) + V(w,u,v) + \tilde{V}(y_u,y_v,y_w)
     - \tilde{V}(y_w,y_u,y_v)]_{w\to 0,\ v\to 1-u} = 0\,.
\label{collinear}
\ee
In fact the parity-odd function $\tilde{V}$ drops out of this constraint.
The reason is that the collinear regime can be approached from the surface
$\Delta(u,v,w)=0$ (see \eqn{zpmDeltadef}), and all parity-odd functions
should vanish on this surface.

The final constraint we will need comes from the predicted OPE behaviour
of the ratio function~\cite{Sever:2011da}. The general philosophy that
an operator product expansion governs the form of the amplitudes
comes from the relation of amplitudes to light-like Wilson
loops. Light-like
Wilson loops can be expanded around a collinear limit and the fluctuations
can be described by operator insertions inside the Wilson
loop~\cite{Alday:2010ku,Gaiotto:2010fk,Gaiotto:2011dt}. By
extending this philosophy~\cite{Sever:2011da} to supersymmetrised Wilson
loops~\cite{CaronHuot:2010ek,Mason:2010yk} (or equivalently
correlation functions~\cite{Eden:2011yp}) one can avoid questions about
giving a precise Lagrangian description of the object under study. In this
sense the OPE can be used to justify the existence of a supersymmetrised
object dual to non-MHV amplitudes.

The analysis of ref.~\cite{Sever:2011da} allows one to choose various
components of the ratio function. Let us consider the component
proportional to $\chi_2 \chi_3 \chi_5\chi_6$. The only term in
\eqn{PVform} that contributes to this component is the first one,
\be
\cP_{\rm NMHV}^{(2356)} = \frac{1}{(2356)} V(u,v,w)\,.
\label{fullcomponent}
\ee
In order to examine the OPE, we follow ref.~\cite{Sever:2011da}
and choose coordinates $(\tau,\sigma,\phi)$ by fixing a conformal frame
where
\be
\frac{1}{(2356)} = \frac{1}{4 (\cosh \sigma \cosh \tau + \cos \phi)}
= \frac{\sqrt{uvw}}{2(1-v)} \,,
\label{prefactor2356}
\ee
and the three cross-ratios are given by
\be
u = \frac{e^\sigma \sinh\tau \tanh\tau}
   {2 (\cosh\sigma \cosh\tau + \cos\phi)}\,, 
\quad v=\frac{1}{\cosh^2\tau}\,, 
\quad w= \frac{e^{-\sigma} \sinh\tau \tanh\tau}
{2 (\cosh\sigma \cosh\tau + \cos\phi)}\,.
\label{uvwtausigmaphi}
\ee

Extrapolating the results of ref.~\cite{Sever:2011da} to two loops,
the OPE predicts the leading (double) discontinuity of the $(2356)$
component of the ratio function to be,
\begin{eqnarray}
\Delta_v \Delta_v \cP_{\rm NMHV}^{(2356)} &\propto&  
\sum_{m=-\infty}^{\infty}  \int_{-\infty}^{\infty} 
\frac{dp}{2 \pi} \, e^{i m \phi- i p \sigma} \, \mathcal{C}_{m}^{(2356)}(p) 
\, \mathcal{F}_{|m|+1,p}^{(2356)}(\tau) \, [ \gamma_{1+|m|}(p) ]^2 \,,
\label{OPEdoubledisc}
\end{eqnarray}
where
\begin{align}
\mathcal{F}_{E,p}^{(2356)}(\tau) &= {\rm sech}^E \tau \,\, 
{}_2F_1\Bigl[\tfrac{1}{2} (E-ip),\tfrac{1}{2}(E+ip);E;{\rm sech}^2 \tau\Bigr],
\\
\mathcal{C}_{m}^{(2356)}(p) &= \tfrac{1}{4} (-1)^m 
B\Bigl[\tfrac{1}{2}(|m|+1+ip),\tfrac{1}{2}(|m|+1-ip)\Bigr]\,,\\
\gamma_{1+|m|}(p) &= \phantom{\Bigl[\Bigr]} \!\!\!\!\! 
   \psi \Bigl(\tfrac{1}{2}(1+|m|+ip)\Bigr)
 + \psi \Bigl(\tfrac{1}{2}(1+|m|-ip)\Bigr) - 2 \psi\bigl( 1 \bigr)\,.
\end{align}
Here ${}_{2}F_{1}$ is the hypergeometric function,
$B(\alpha,\beta) = \Gamma(\alpha) \Gamma(\beta)/\Gamma(\alpha + \beta)$
is the Euler beta function, and $\psi$ is the logarithmic derivative of
the $\Gamma$ function.


\section{Ansatz for the symbol of the two-loop ratio function}
\label{sect-ansatz}
\setcounter {equation} {0}

In order to make a plausible ansatz for the ratio function at two loops we
assume that the functions $V^{(2)}(u,v,w)$ and $\tilde{V}^{(2)}(u,v,w)$
are {\sl pure} functions of $u$, $v$ and $w$, {\it i.e.}~iterated
integrals or multi-dimensional polylogarithms of degree four. Moreover, we
make an ansatz for the symbols of $V^{(2)}$ and $\tilde{V}^{(2)}$,
requiring that their entries are drawn from the following set of
nine elements,
\be
\{ u,v,w,1-u,1-v,1-w,y_u,y_v,y_w\}\,.
\label{letters}
\ee
We summarise some background material on pure functions and symbols in
appendix \ref{app-symbols}. We recall that the $y$ variables invert under
parity. The parity-even function $V$ should have a symbol which contains
only terms with an even number of $y$ entries. Likewise the symbol of the
parity-odd function $\tilde{V}$ should contain only terms with an odd
number of $y$ entries.  The ansatz for the symbol entries is the same 
as the one used recently for the three-loop remainder
function~\cite{Dixon:2011pw} (after omitting restrictions on the
final entry of the symbol).  It is
consistent with every known function appearing in the six-point amplitudes
of planar $\cN=4$ super Yang-Mills theory, in particular the simple
analytic form of the two-loop remainder function found in
ref.~\cite{Goncharov:2010jf}.  It is also consistent with the results
for explicitly known loop integrals appearing in such amplitudes,
see refs.~\cite{Drummond:2010cz,Dixon:2011ng,DelDuca:2011ne}.
In the ensuing analysis we will find many
strong consistency checks on our ansatz.

Let us pause to note that our assumption that the relevant functions are
pure functions of a particular degree equal to twice the loop order is by
no means an innocent one. Although it is true that such general
polylogarithmic functions generically show up in amplitudes in
four-dimensional quantum field theories, it is certainly not true in
general that they always appear with a uniform degree dependent on the
loop order. In QCD, for example, the degrees appearing range from twice the
loop order to zero, and the transcendental functions typically appear with
non-trivial algebraic prefactors. In fact the observed
behaviour of having maximal degree only is limited to $\cN=4$ super
Yang-Mills theory, and the most evidence is for the planar sector.
This behavior is the generalization, to non-trivial functions of the
kinematics, of the maximal degree of transcendentality for harmonic
sums that has been observed in the anomalous dimensions of
gauge-invariant local operators~\cite{Kotikov:2004er}.

The symbols we construct from the set of letters~(\ref{letters}) should
obey certain restrictions. They should be integrable; that is, they should
actually be symbols of functions. The initial entries of the symbol should
be drawn only from the set $\{u,v,w\}$, because the leading entry
determines the locations of branch points of the function in question,
and branch  integrable symbols of degree 4 for $V$, and 2 for
$\tilde{V}$, obeying the initial entry condition as well as the
symmetry conditions~(\ref{sym}). The spurious pole
conditions~(\ref{spurious}) provide 14 constraints and the collinear
conditions~(\ref{collinear}) provide 14 more, leaving 15 free parameters
at this stage.  In order to impose the
constraints from the leading discontinuity predicted by the OPE, we
use the fact that the sum~(\ref{OPEdoubledisc}) is annihilated by the
following differential operator~\cite{Sever:2011da},
\begin{eqnarray}
 {\mathcal D} &=& \partial_{\tau}^2 + 2 \coth(2 \tau) \, \partial_\tau
 + {\rm sech}^2\tau \, \partial_\sigma^2 + \partial_\phi^2 + 1\,.
\end{eqnarray}
In the $u,v,w$ variables this differential operator is given by
\begin{eqnarray}
 {\mathcal D}  &=& \tfrac{1}{2} ( \mathcal{D}_{+} + \mathcal{D}_{-}) + 1\,,
\label{D_defn}
\end{eqnarray}
where 
\begin{align}
\mathcal{D}_\pm = \frac{4}{1-v}\Big[ 
&- z_\pm u \partial_u -(1-v)v \partial_v - z_\pm w \partial_w \notag \\
&+ (1-u)v u\partial_u u \partial_u 
 + (1-v)^2 v \partial_v v \partial_v
 + (1-w)vw\partial_w w \partial_w \notag \\
&+ (-1+u-v+w)\bigl((1-v)u\partial_uv\partial_v
 - vu \partial_u w \partial_w + (1-v)v\partial_v w \partial_w\bigr)\Bigr]\,.
\label{Dpmuvw}
\end{align}

Imposing that the double discontinuity~(\ref{fullcomponent})
is annihilated by the operator $\mathcal{D}$
gives 5 further conditions, leaving 10 free parameters in
the symbol. One of these parameters, denoted by $\alpha_X$ below, is just
the overall normalisation of the symbol of the double discontinuity, which
is non-zero and convention-dependent. In the following we provide
functions with physical branch cuts which represent the symbol.
We find that the solution has the form,
\be \label{ansatzOPEimposed}
\mathcal{S}(V)\ =\ 
\alpha_X \, \mathcal{S}(V_X)
+ \sum_{i=1}^9 \alpha_i \, \mathcal{S}(f_i), \qquad 
\mathcal{S}(\tilde{V}) 
\ =\ \alpha_X \, \mathcal{S}(\tilde{V}_X)
 + \alpha_8 \, \mathcal{S}(\tilde{f})\,,
\ee
where $\alpha_X$ and $\alpha_1$ through $\alpha_9$ are the constant
free parameters, and the quantities $V$, $f_i$, $\tilde{V}_X$ and
$\tilde{f}$ will be defined below.\footnote{%
In section~\ref{sect-analytic} we will fix the ten parameters using 
an analytical computation for particular kinematics.
We have compared the symbols~(\ref{ansatzOPEimposed}) for $V$ and
$\tilde{V}$ with all parameters fixed to an independent computation of
these symbols from a formulation of the super Wilson
loop~\cite{SimonPrivate}; the results agree precisely.}

The double discontinuities of the functions appearing in
\eqn{ansatzOPEimposed} obey
\bea
\mathcal{S}(\Delta_v \Delta_v V_X) &=& 
2 \, \alpha_X \Bigl[
 u \otimes (1-u) \, + \, u \otimes u
\, + \, w \otimes (1-w) \, + \, w \otimes w
\nonumber\\  &&\hskip 0.7cm \null
+ 2 \Bigl( 
u \otimes w + w \otimes u 
- uw \otimes (1-v) - (1-v) \otimes uw 
\nonumber\\  &&\hskip 1.0cm \null
+ (1-v) \otimes (1-v) \Bigr) \Bigr] \,,
\label{doublevVX} \\
\mathcal{S}(\Delta_v \Delta_v f_i) &=& 0\,,
\label{doublevfi} \\
\mathcal{S}(\Delta_v \Delta_v \tilde{V}_X) &=& 
\mathcal{S}(\Delta_v \Delta_v \tilde{f})\ =\ 0\,.
\label{doublevtildeVXf}
\eea
Consistency with the spurious pole condition~(\ref{spurious})
forces the odd part $\tilde{V}$ to be non-zero, given that $\alpha_X$
is non-zero.  The odd part contains no ambiguity at the level of the symbol
(or beyond it), once we fix the even part, particularly the two
parameters $\alpha_X$ and $\alpha_8$.  

The symbol of the double discontinuity of $V$ and $\tilde{V}$ is
entirely controlled by $V_X$, through \eqn{doublevVX}.
We can find a function compatible with this symbol, and compare it
to \eqn{OPEdoubledisc} to fix the $\zeta_2$ terms.  We find that
\begin{eqnarray}
\Delta_v \Delta_v \cP_{\rm NMHV}^{(2356)} 
&\propto& \frac{1}{(2356)} 
\bigg[  \log^2 u +\log^2 w + 4 \, \log u \, \log w
+ 2 \, \log^2(1-v)\nonumber \\
&&\null - 4 \log (u w) \, \log (1-v)
- 2 \, \Bigl( \text{Li}_2(1-u) + \text{Li}_2(1-w) - 2 \, \zeta_2 \Bigr)
 \bigg] \,.
\end{eqnarray}

In order to present the symbols appearing in \eqn{ansatzOPEimposed}
explicitly and compactly, it is very useful to employ harmonic
polylogarithms~\cite{Remiddi:1999ew,Gehrmann:1999as,Gehrmann:2001pz}.
This presentation simultaneously accomplishes the following step,
of turning the symbols into functions, up to certain beyond-the-symbol
ambiguities.  The functions we will present are of degree at most four,
and almost all of them can be represented in terms of classical
${\rm Li}_n$ functions.  Thus the use of harmonic polylogarithms may seem
unnecessarily complicated.  However, it is a very useful way to represent,
at any degree, a symbol only involving the letters $\{u,v,w,1-u,1-v,1-w\}$,
whereas ${\rm Li}_n$ functions are often insufficient beyond degree four.

Harmonic polylogarithms are single-variable functions defined by
iterated integration.  It is very simple to write down their
symbols. We use harmonic polylogarithms with labels (weight-vector entries)
``0" and ``1" only.  The symbol of a harmonic polylogarithm of argument $x$
is obtained by reversing the list of labels and replacing all ``0"
entries by $x$ and all ``1" entries by $1-x$.  Finally, one multiplies
by $(-1)^n$ where $n$ is the number of ``1" entries. For example, the
symbol of $H_{0,0,0,1}(x) = {\rm Li}_4(x)$ is
$-\ (1-x)\otimes x\otimes x\otimes x$, and the symbol of
$H_{0,1,0,1}(x)$ is $(1-x)\otimes x \otimes (1-x) \otimes x$\,.

We also use the common convention of shortening the label list by deleting
each ``0" entry, while increasing by one the value of the first non-zero
entry to its right, so that, for example,
$H_{0,0,0,1}(x) = H_4(x) = {\rm Li}_4(x)$
and $H_{0,0,1,1}(x) = H_{3,1}(x)$.
Apart from the logarithm function, we take all arguments of the
harmonic polylogarithms to be $(1-u)$, $(1-v)$ or $(1-w)$.  This
representation guarantees that the functions we are
using to represent the symbol do not have any branch cut originating
from an unphysical point. We then compactify the notation further by
writing $H_{2,2}(1-x) = H^x_{2,2}$, and so on.
Finally, we recall that the symbol of a product of two functions is
given by the shuffle product of the two symbols.

With this notation we can immediately write down a function which has the
symbol, $\mathcal{S}(V_X)$, of the part of $V$ with 
non-zero double discontinuity, {\it i.e.}~the part fixed by the OPE,
\begin{align}
V_X = \Bigl\{ &4 H^u_{3, 1} +  \log u \, ( H^v_{3} + 2 H^u_{2, 1}  - 
 5 H^v_{2, 1} + 6 H_{2, 1}^w +\tfrac{3}{2} H^u_{2}  \log w) + 
 \log^2 u \, (H^u_{2}  - 3 H^w_{2})  \notag \\
 &+  \log v \, [ H^u_{3} - 3 H^u_{2, 1}
  - \tfrac{1}{2} \log u \, ( H^u_{2} +  H^v_{2} )
  - \tfrac{3}{2} \log^2 u \log w ]
  +  \log^2 v \, ( - H^u_{2} + \tfrac{1}{2} \log^2 u )  \notag \\
 & + (u \leftrightarrow w) \Bigr\}
 + 4 H^v_{3, 1} 
 +  2 \log u \log^2 v \log w - \tfrac{1}{2} \log^2 u \log^2 w \,.
\end{align}
We can similarly write down functions with the correct symbols
for the first seven ambiguities in the even part (the double $v$
discontinuity of each function vanishes):
\begin{align}
f_1 = \phantom{[}& \!\! H^u_{2} H^w_{2}\,, \notag \\
& \notag \\
f_2 = [&-\log u \, ( H^w_{3} + H^w_{2, 1} + H^u_{2} \log w )
  - \log^2 u \, ( H^w_{2}  + \tfrac{1}{2}  \log v \log w )
 + (u \leftrightarrow w)]  \,,  \notag \\
&\notag \displaybreak[0]\\
f_3= [& - H^w_{2} \log u \log v + (u \leftrightarrow w)]
 + H^v_{2} \log u \log w\,, \notag \\
& \notag \displaybreak[0]\\
f_4 = [& - H^u_{2} \log u \log w -\log^2 u \, ( 2 H^w_{2}  + \log v \log w)
 + (u \leftrightarrow w)] - H^v_{2} \log u \log w\,,  \notag \\
& \notag \displaybreak[0]\\
f_5 = [&H^u_{2} H^v_{2} + H^u_{2, 2}
 + \log u \, ( 2 H^v_{2, 1}  - 2 H^w_{2, 1} -  H_{2}^u  \log w )
 + \log v \, ( 2 H^u_{2, 1} +  \log u \, ( H^u_{2} + H^v_{2} ) ) \notag \\
 &+ (u \leftrightarrow w)]  + H^v_{2, 2} \,, \notag \\
 &\notag \displaybreak[0]\\
f_6 = [&-2 H^u_{2}H^v_{2} - 2 H^u_{2, 2} -4 H^u_{3, 1}
 + \log u \, ( - 2 H^u_{3}  - 2 H^v_{2, 1}   +  2 H^w_{2, 1}
           + H^u_{2}  \log w ) \notag \\
&-\log v \, ( 2 H^u_{2, 1}  + \log u \, ( H^u_{2} + H^v_{2} ) )
 + (u \leftrightarrow w)]  
 - 2 H^v_{2, 2} - 4 H^v_{3, 1}  - 2 H^v_{3} \log v \,,   \notag \\
 &\notag \displaybreak[0]\\
f_7 = [&-3 H^u_{4} - 3 H^u_{2, 1, 1}
  + \log u \, ( H^v_{3} - 2 H^u_{2, 1} + H^w_{2, 1} + H^u_{2}  \log w )
+ \log^2 u \, ( - \tfrac{1}{2} H^u_{2}  + H^w_{2}) \notag \\
& + \log v \, ( H^u_{3}  + \tfrac{1}{2} \log^2 u \log w )
   + (u \leftrightarrow w)] 
 - 3 H^v_{4}  - 3 H^v_{2, 1, 1}  
 - 2 H^v_{2, 1} \log v - \tfrac{1}{2} H^v_{2} \log^2 v \,. \notag
\end{align}

For the even part there remain two more ambiguities whose symbols
cannot be expressed in terms of those of the single-variable harmonic
polylogarithms with the arguments we have been using,
\begin{align}
f_8 =  [&-H^u_{2} H^v_{2}  - 2 H^u_{4} + H^u_{2, 2} - 4 H^u_{3, 1}
    + 6 H^u_{2, 1, 1} 
+ \log u \, ( - H^u_{3} - H^v_{3} + H^w_{3}  + 2 H^u_{2, 1})  \notag \\
  &+  H^w_{2} \log^2 u + \log v \, ( H^u_{3} - H^w_{2} \log u )
 + (u \leftrightarrow w)] \notag \\
  - & H^v_{2, 2} - 2 H^v_{3, 1} + H^v_{2} \log u \log w 
  + \tfrac{1}{2} \log^2 u \log^2 w  - H^v_{3} \log v
  - 2 \Omega^{(2)}(w,u,v) \,, \notag \\
  &\notag \\
f_9 = \phantom{[}&\mathcal{R}_6^{(2)}(u,v,w)\,. \notag
\end{align} 
Here $\mathcal{R}_6^{(2)}$ stands for the two-loop remainder function,
whose symbol is known~\cite{Goncharov:2010jf}. The appearance of
the two-loop remainder function as an ambiguity should not be
surprising. It is a function with physical branch cuts, which vanishes in
the collinear limit.  Also, it is totally cyclic and hence automatically
satisfies the spurious pole condition on its own. Furthermore, it has
vanishing double discontinuities and hence drops out from the
leading-discontinuity OPE criterion~(\ref{OPEdoubledisc}).
It is known~\cite{Goncharov:2010jf} that
$\mathcal{R}_6^{(2)}$ can in fact be expressed in terms of single-variable
classical polylogarithms.  However, to do so one must use arguments
involving square roots of polynomials of the cross-ratios. 

The other quantity not given in terms of the harmonic polylogarithms, which
enters $f_8$, is the integral $\Omega^{(2)}$.  In fact its symbol can also 
be recognised from other
considerations~\cite{Drummond:2010cz,Dixon:2011ng}, as we will discuss in
the next section.  The symbol of $\Omega^{(2)}$ is,
\be
\mathcal{S}(\Omega^{(2)}(u,v,w))
= -\frac{1}{2} \Bigl[ \mathcal{S}(q_{\phi}) \otimes \phi
                     + \mathcal{S}(q_r) \otimes r
                     + \mathcal{S}(\tilde{\Phi}_6)\otimes y_u y_v \Bigr]\,,
\label{S(O2)}
\ee
where
\be
\phi = \frac{u v}{(1-u)(1-v)}, \qquad r = \frac{u(1-v)}{v(1-u)} \,.
\ee
Here $\tilde{\Phi}_6$ is the one-loop six-dimensional hexagon
function~\cite{Dixon:2011ng,DelDuca:2011ne}, whose symbol is given
explicitly in terms of the letters of our ansatz~\cite{Dixon:2011ng},
\be
\mathcal{S}(\tilde{\Phi}_6)
= - \mathcal{S}\bigl( \Omega^{(1)}(u,v,w)\bigr) \otimes y_w
+ {\rm cyclic}, 
\ee
where $\Omega^{(1)}$ is a finite, four-dimensional one-loop hexagon
integral~\cite{ArkaniHamed:2010kv,Drummond:2010cz},
\be
\Omega^{(1)}(u,v,w) = \log u \log v
 + {\rm Li}_2(1-u) + {\rm Li}_2(1-v) + {\rm Li}_2(1-w) - 2 \zeta_{2}\,.
\ee
The other degree 3 symbols above can be represented by harmonic
polylogarithms as follows,
\begin{align}
q_\phi\ &=\ [-H^u_{3} - H^u_{2, 1} - H^v_{2} \log u
 - \tfrac{1}{2} \log^2 u \log v + H^u_{2} \log w
 + (u \leftrightarrow v)] \notag \\
&\quad + 2 H^w_{2, 1}  + H^w_{2}  \log w +  \log u \log v \log w\,, \notag \\
q_r\ &=\ [-H^u_{3} + H^u_{2, 1} + H^u_{2} \log u + H^w_{2} \log u
 + \tfrac{1}{2} \log^2 u \log v - (u \leftrightarrow v)]   \,.
\label{qphiqr}
\end{align}

For the symbol of the parity-odd function $\tilde{V}$, we find that the
part fixed by the OPE (acting in conjunction with the spurious-pole
constraint~(\ref{spurious})), $\mathcal{S}(\tilde{V}_X)$, coincides
with the symbol of
\be
\tilde{V}_X\ =\ \tilde{\Phi}_6 \log \Bigl(\frac{u}{w}\Bigr)\,.
\label{tV_X}
\ee
For the odd part of the ambiguity associated with $\alpha_8$, we have
\be
\mathcal{S}(\tilde{f})\ =\ \mathcal{S}(\tilde{f}_u) \otimes y_u
 + \mathcal{S}(\tilde{f}_v )\otimes y_v
 + \mathcal{S}(\tilde{f}_w) \otimes y_w
 - \mathcal{S}(\tilde{\Phi}_6) \otimes \frac{1-u}{1-w}\,,
\ee
where the functions $\tilde{f}_u, \tilde{f}_v, \tilde{f}_w$ are given by,
\begin{align}
\tilde{f}_u &= [2 H^u_{3} - H^u_{2} \log v - (u \leftrightarrow w)]
  - 2 H^v_{2, 1}  - H^v_{2} \log v\,, \notag\\
\tilde{f}_v &= [2 H^u_{3} - 2 H^u_{2, 1} - H^u_{2} \log u
 - H^v_{2} \log u + H^w_{2} \log u - (u \leftrightarrow w)] \,, \notag\\
\tilde{f}_w &= [2 H^u_{3} - H^u_{2} \log v - (u \leftrightarrow w)]
  + 2 H^v_{2, 1}  + H^v_{2} \log v\,.
\end{align}

We emphasise again that the formulas presented in this section are meant
to represent the symbols of the functions involved. For some of the
relevant symbols ($\mathcal{S}(V_X)$ and
$\mathcal{S}(f_1)$ through $\mathcal{S}(f_7)$) we were able to trivially
write down actual functions
which represent those symbols in terms of single-variable harmonic
polylogarithms with arguments $1-x$, where $x$ is one of the
cross-ratios. For two others ($\mathcal{S}(f_9)$ and
$\mathcal{S}(\tilde{V}_X)$) we recognised them as involving symbols of
functions we already know, namely the two-loop remainder function and the
one-loop six-dimensional hexagon integral.  In order to write down actual
functions for $V$ and $\tilde{V}$ there are two issues to
address. Firstly, we must give functions which represent the symbols
$\mathcal{S}(\Omega^{(2)}(w,u,v))$ and $\mathcal{S}(\tilde{f})$.
Secondly, we must include all possible terms which have vanishing symbol
and which are therefore insensitive to the analysis we have presented so
far.  We address these two issues in the next two sections.


\section{Digression on integral representations
for $\Omega^{(2)}$ and $\tilde{f}$}
\label{sect-symboltofunction1}
\setcounter {equation} {0}

Here we will present integral formulas to define the functions
$\Omega^{(2)}$ and $\tilde{f}$ whose symbols are given in the
previous section.  We also present a new representation of the
two-loop remainder function, based on the integral $\Omega^{(2)}$.  
This section is more technical, and could therefore be skipped on a
first reading.

\subsection{Integral representations for $\Omega^{(2)}$}

We start with the finite double-pentagon integral
$\Omega^{(2)}(u,v,w)$~\cite{ArkaniHamed:2010kv}.
Let us take a derivative with respect to $w$. 
The only contributing term from the symbol~(\ref{S(O2)}) is the last one,
so we see that the symbol $\mathcal{S}(\Omega^{(2)})$ is consistent with
the differential equation,
\be
\partial_w \Omega^{(2)}(u,v,w)
= -\frac{\tilde{\Phi}_6}{2} \, \partial_w \log (y_u y_v)
= -\frac{\tilde{\Phi}_6}{\sqrt{\Delta}} \,.
\label{dwO2=Phi6}
\ee
We recognise here the differential equation~\cite{Dixon:2011ng}
relating the two-loop, finite double-pentagon integral $\Omega^{(2)}$ to
the massless, one-loop, six-dimensional hexagon function
$\tilde{\Phi}_6$. 

The relation~(\ref{dwO2=Phi6}) can be used to write an
integral formula for $\Omega^{(2)}$,
\be
\Omega^{(2)}(u,v,w)\ =\ - \int_0^w \frac{dt}{\sqrt{\Delta(u,v,t)}}
 \, \tilde{\Phi}_6(u,v,t)\, + \Omega^{(2)}(u,v,0)\,.
\label{Omega2fromtPhi6}
\ee
The relevant boundary condition is
$\Omega^{(2)}(u,v,0) = \Psi^{(2)}(u,v)$, 
where $\Psi^{(2)}$ is the two-loop pentaladder function
found in ref.~\cite{Drummond:2010cz}. The boundary behaviour at $w=0$ was
tested numerically from the Mellin-Barnes representation for
$\Omega^{(2)}$~\cite{Drummond:2010cz}. The symbol~(\ref{S(O2)}) reduces to
the symbol of $\Psi^{(2)}$ at $w=0$. It is the unique symbol within our
ansatz, built from the letters in \eqn{letters}, that obeys~\eqn{dwO2=Phi6}
and the $w=0$ boundary condition.  The integral in \eqn{Omega2fromtPhi6}
is well-defined and real in the Euclidean region, {\it i.e.}~the positive
octant in which $u,v,w$ are all positive, because the integrand
$\Phi_6 \equiv \tilde{\Phi}_6/\sqrt{\Delta}$ is well-defined and real there,
and $\Phi_6$ is well-behaved even where $\Delta$ vanishes~\cite{Dixon:2011ng}.

As discussed in ref.~\cite{Dixon:2011ng}, the first-order differential
equation~(\ref{dwO2=Phi6}) can be obtained from the second-order equation
of ref.~\cite{Drummond:2010cz} for the double-pentagon integral,
which can be written as
\be
w\partial_w \Bigl[ - u(1-u)\partial_u - v(1-v)\partial_v
                   + (1-u-v)(1-w)\partial_w \Bigr]
 \Omega^{(2)}(u,v,w) =  \Omega^{(1)}(u,v,w)\,.
\ee
Because the second-order operator naturally factorises into two first-order
operators, we can integrate up to $\Omega^{(2)}$ in two steps.  This
procedure will yield another one-dimensional integral relation for
$\Omega^{(2)}$.  We define
\begin{align}
Q_\phi(u,v,w) \equiv
\Bigl[ -u(1-u)\partial_u - v(1-v)\partial_v 
+(1-u-v)(1-w)\partial_w \Bigr] \Omega^{(2)}(u,v,w) \,,
\label{dQphi=O2}
\end{align}
so that
\be
w\partial_w Q_\phi(u,v,w) = \Omega^{(1)}(u,v,w)\,.
\label{dQphi}
\ee
The above formula can be used to define the function $Q_\phi$,
\begin{align}\label{Qphi1}
Q_\phi(u,v,w)\ =& \  
2 \, \biggl[ {\rm Li}_3(1-w) + {\rm Li}_3\left(1-\frac{1}{w}\right) \biggr]
\\ \nonumber &\null
+ \ln w \, \Bigl[ - {\rm Li}_2(1-w) 
+ {\rm Li}_2(1-u) + {\rm Li}_2(1-v) + \ln u \, \ln v 
- 2 \, \zeta_2 \Bigr]
\\ \nonumber &\null
- \frac{1}{3} \ln^3 w
- 2 \, {\rm Li}_3(1-u) - {\rm Li}_3\left(1-\frac{1}{u}\right) 
- 2 \, {\rm Li}_3(1-v) - {\rm Li}_3\left(1-\frac{1}{v}\right)
\\ \nonumber &\null
+ \ln\left(\frac{u}{v}\right) 
\Bigl[ {\rm Li}_2(1-u) - {\rm Li}_2(1-v) \Bigr]
+ \frac{1}{6} \, \ln^3 u + \frac{1}{6} \, \ln^3 v
\\ \nonumber &\null
- \frac{1}{2} \, \ln u \, \ln v \, \ln(uv) \,. 
\end{align}
This function obeys \eqn{dQphi} and has a symbol
coinciding with that of $q_\phi$ from \eqn{qphiqr}. It also obeys
$Q_\phi(1,1,1)=0$.
In principle, \eqn{dQphi} allows one to add beyond-the-symbol terms
to $Q_\phi$ that are proportional to $\zeta_2 \log( u v )$, and to
$\zeta_{3}$. We verified numerically that these terms are absent.
The function $Q_\phi$ is manifestly real in the positive octant.

Given the function $Q_\phi$, we can integrate \eqn{dQphi=O2} to obtain
$\Omega^{(2)}$.  We first note that the relevant operator
becomes very simple in the $(y_u,y_v,y_w)$ variables,
\be
- u(1-u)\partial_u - v(1-v)\partial_v + (1-u-v)(1-w)\partial_w
= \frac{(1-y_w)(1-y_uy_vy_w)}{1-y_u y_v} \, \partial_{y_w}\,.
\ee
Inserting this relation into \eqn{dQphi=O2}, we find an alternative
integral formula for $\Omega^{(2)}$ in terms of the $y$ variables,
\be\label{Omeganum}
\Omega^{(2)}(u,v,w)\ =\ - 6 \, \zeta_{4}
+ \int_{\frac{1}{y_u y_v}}^{y_w} 
\frac{dt}{1-t} \frac{1-y_u y_v}{1-y_u y_v t} \, \hat{Q}_\phi(y_u,y_v,t)
\,.
\ee
Here we use the notation
$\hat{Q}_\phi(y_u,y_v,y_w)
= Q_\phi(u(y_u,y_v,y_w),v(y_u,y_v,y_w),w(y_u,y_v,y_w))$.
Note that this integral is well-defined at the lower limit of
integration, for the following reason:  Whenever the product of
the $y$ variables is unity, $y_u y_v y_w = 1$, we see from
\eqn{u1mufromy} that the cross-ratios collapse to the point
$(u,v,w)=(1,1,1)$, and at that point $Q_\phi$ vanishes, 
$Q_\phi(1,1,1)=0$.
\Eqn{Omeganum} can be applied straightforwardly in the
$y$ variables for $y_w > 1$ and $y_u y_v < 1$, and also
for $y_w < 1$ and $y_u y_v > 1$.   (In other regions, the
vicinity of $t=1$ makes a direct integration problematic.)

It can be more convenient to map the integral~(\ref{Omeganum})
back to the $(u,v,w)$ space.  This mapping avoids problems related to 
the variables $(y_u,y_v,y_w)$ becoming complex when $\Delta$ is negative.
To do this mapping, we first define
\begin{eqnarray}
r &=& \frac{u(1-v)}{v(1-u)}
  \ =\ \frac{y_u(1-y_v)^2}{y_v(1-y_u)^2} \,, \label{rdef}\\
s &=& \frac{u(1-u)v(1-v)}{(1-w)^2}
  \ =\ \frac{y_u(1-y_u)^2 \, y_v(1-y_v)^2}{(1-y_u y_v)^4} \,, \label{sdef}\\
t &=& \frac{1-w}{uv} 
  \ =\ \frac{(1-y_u y_v)^2 \, (1-y_u y_v y_w)}
       {y_u(1-y_u) \, y_v(1-y_v) \, (1-y_w)} \,. \label{tdef}
\end{eqnarray}
Notice that $r(y_u,y_v,y_w)$ and $s(y_u,y_v,y_w)$ are actually
independent of $y_w$.  Therefore the curve of integration in the
integral~(\ref{Omeganum}) from $(1,1,1)$ to $(u,v,w)$, which has
constant $y_u$ and $y_v$, should have a 
constant value of $r$ and $s$ along it, while $t$ varies.
Also,
\begin{equation}
\frac{d(\ln t)}{dy_w} = \frac{1-y_u y_v}{(1-y_w)(1-y_u y_v y_w)} \,,
\label{Jactyw}
\end{equation}
so that the measure in \eqn{Omeganum} is just $d \ln t$.

Let $(u_t,v_t,w_t)$ be the values of $(u,v,w)$ along the curve
from $(1,1,1)$ to $(u,v,w)$.  We solve the two constraints, that
$r$ and $s$ are constant along the curve, {\it i.e.}
\begin{eqnarray}
\frac{u_t(1-v_t)}{v_t(1-u_t)} &=& \frac{u(1-v)}{v(1-u)}
\,, \label{rconst}\\
\frac{u_t(1-u_t)v_t(1-v_t)}{(1-w_t)^2}
 &=& \frac{u(1-u)v(1-v)}{(1-w)^2} \,, \label{sconst}
\end{eqnarray}
for $v_t$ and $w_t$ in terms of $u_t$, obtaining,
\begin{eqnarray}
v_t &=& \frac{(1-u) \, v \, u_t }{ u \, (1-v) + (v-u) \, u_t } \,,
\label{vtdef}\\
w_t &=& 1 
- \frac{ (1-w) \, u_t \, (1-u_t) }{ u \, (1-v) + (v-u) \, u_t } \,.
\label{wtdef}
\end{eqnarray}
Inserting these expressions into $d\ln t = d\ln[(1-w_t)/u_t/v_t]$,
we have
\begin{equation}
\frac{d(\ln t)}{du_t} = \frac{1}{u_t(u_t-1)} \,,
\label{Jactut}
\end{equation}
which enables us to use $u_t$ as the final integration parameter,
\begin{eqnarray}\label{Omeganumnew}
\Omega^{(2)}(u,v,w) = - 6 \, \zeta_4
 + \int_1^u \frac{du_t}{u_t(u_t-1)} \, Q_\phi(u_t,v_t,w_t) \,.
\end{eqnarray}
Using this formula, with $Q_\phi$ from \eqn{Qphi1}, 
for which the polylogarithms are all rational functions of
the cross ratios, it is easy to rapidly get high-accuracy values
for $\Omega^{(2)}$.  For example, we find
\begin{eqnarray}
\Omega^{(2)}(\tfrac{28}{17},\tfrac{16}{5},\tfrac{112}{85})
&=& -5.273317108708980008 \,, 
\label{Omega2numKRV1}\\
\Omega^{(2)}(\tfrac{16}{5},\tfrac{112}{85},\tfrac{28}{17})
&=& -6.221018431345742955 \,,
\label{Omega2numKRV2}\\
\Omega^{(2)}(\tfrac{112}{85},\tfrac{28}{17},\tfrac{16}{5})
&=& -9.962051212650647413 \,,
\label{Omega2numKRV3}
\end{eqnarray}
in general agreement with the numbers obtained at these points
using a Mellin-Barnes representation for the loop integral.

\subsection{A new representation of the two-loop remainder function}

Now that we have obtained representations of the function $\Omega^{(2)}$,
we note that the two-loop remainder function can be written in terms of this
function, together with functions with purely rational ($y$-independent)
symbols.  Specifically, we have
\be\label{remaindernewrepresentation}
\mathcal{R}_6^{(2)}(u,v,w) = \frac{1}{4} \Bigl[
\Omega^{(2)}(u,v,w) + \Omega^{(2)}(v,w,u) + \Omega^{(2)}(w,u,v) \Bigr]
+ \mathcal{R}_{6,{\rm rat}}^{(2)}\,.
\ee
The piece with a rational symbol is defined as
\be\label{R62ratdef}
\mathcal{R}_{6, {\rm rat}}^{(2)} = 
- \frac{1}{2} \biggl[ 
\frac{1}{4}\Bigl( {\rm Li}_2(1-1/u) + {\rm Li}_2(1-1/v)
                  + {\rm Li}_2(1-1/w)\Bigr)^2
+ r(u) + r(v) + r(w) - \zeta_{4} \biggr]\,,
\ee
with
\begin{align}\label{rudef}
r(u) = & - {\rm Li}_4(u) - {\rm Li}_4(1-u) + {\rm Li}_4(1-1/u)
- \log u \, {\rm Li}_3(1-1/u) - \frac{1}{6} \, \log^3 u \, \log(1-u) \notag \\
& + \frac{1}{4} \Bigl({\rm Li}_2(1-1/u) \Bigr)^2 + \frac{1}{12} \log^4 u
+ \zeta_{2} \Bigl({\rm Li}_2(1-u)+\log^2 u \Bigr)+ \zeta_{3} \, \log u \,.
\end{align}
The function $\mathcal{R}_{6, {\rm rat}}^{(2)}$ is real when all
three cross-ratios are positive.  Almost all of the terms in
\eqns{R62ratdef}{rudef} make this manifest term-by-term, because they contain
only logarithms of cross-ratios, or ${\rm Li}_n(x)$ for some argument
$x$ which is less than one.  The one slight exception is the combination
\be
\polylog_4(u) + \frac{1}{6} \, \log^3 u \, \log(1-u)\,.
\label{li4cancelsum}
\ee
It is easy to see that \eqn{li4cancelsum} is real as well, but in this
case the branch cut starting at $u=1$ in each term cancels in the sum.

In one sense, the representation~(\ref{remaindernewrepresentation})
is a step backward from ref.~\cite{Goncharov:2010jf}, because the function
$\Omega^{(2)}(u,v,w)$ cannot be expressed in terms of classical
polylogarithms, whereas $\mathcal{R}_6^{(2)}$ can be.  (The absence of a
classical polylogarithmic representation for $\Omega^{(2)}$ can be seen
from its symbol, using the test described in ref.~\cite{Goncharov:2010jf}.)
However, the appearance of the sum over cyclic permutations of the finite
two-loop double-pentagon integral is natural, and the coefficient of
$\tfrac{1}{4}$ matches the one in the expression for the two-loop MHV
amplitude in ref.~\cite{ArkaniHamed:2010kv}.
The relation~(\ref{remaindernewrepresentation}) between
$\mathcal{R}_6^{(2)}$ and $\Omega^{(2)}$ will be useful for us in the
ensuing NMHV analysis.

\subsection{An integral representation for $\tilde{f}$}

We can obtain in a similar way an integral formula for the parity-odd
function $\tilde{f}$.   Note that we already have a formula, \eqn{tV_X}, 
for the function $\tilde{V}_X = \tilde{\Phi}_6 \ln(u/w)$.
It is useful to observe that the combination
$\tilde{V}_X + \tilde{f}$ has a symbol which can be arranged so that the
final entries are drawn from the list,
\be
\left\{ y_u,y_v,y_w,\frac{u(1-w)}{w(1-u)}\right\} \,.
\ee
In terms of the $y$ variables, the last final entry in the list above is
independent of $y_v$,
\be
\frac{u(1-w)}{w(1-u)} = \frac{y_u (1-y_w)^2}{y_w (1-y_u)^2}\,.
\ee
This fact allows us to obtain the symbol of the logarithmic
derivative with respect to $y_v$, which is independent
of the $y$ variables,
\bea
{\cal S}(\tilde{Z}) &=& 2 \, \biggl\{ u \otimes \frac{u}{1-u} \otimes (1-u)
- \Bigl[ u \otimes \frac{w}{1-u} - v \otimes (1-v) + w \otimes u \Bigr]
   \otimes u
\nonumber\\
&&\hskip0.6cm\null
+ \bigl( u \otimes v + v \otimes u \bigr) \otimes (1-v)
- u \otimes u \otimes w \biggr\} \,.
\label{Ztilde_symb}
\eea
The combination
$\tilde{V}_X + \tilde{f}$ can then be written as an integral of a
function with this symbol,
\be \tilde{V}_X + \tilde{f} = \int_{\frac{1}{y_u y_w}}^{y_v}
\frac{dt}{t} \, \tilde{Z}(y_u,t,y_w) \,.
\label{Vtildeyform}
\ee
Here $\tilde{Z}$ is to be considered as a function of the variables
$(y_u,y_v,y_w)$ for the integration, but it is most simply expressed
in terms of the variables $(u,v,w)$,
\bea
\tilde{Z}(u,v,w) &=& - 2 \, \biggl[ 
    \polylog_3\left(1-\frac{1}{u}\right) 
  - \polylog_3\left(1-\frac{1}{w}\right)
  + \ln\left(\frac{u}{w}\right)
 \Bigl( \polylog_2(1-v) - 2 \, \zeta_2 \Bigr)
\nonumber\\ &&\null\hskip0.7cm 
  - \frac{1}{6} \, \ln^3\left(\frac{u}{w}\right) \biggr]
\,,
\label{Ztilde}
\eea
in which form it is manifestly real in the positive octant.

Using the same trick we used for $\Omega^{(2)}$,
we can rewrite the integral~(\ref{Vtildeyform}) directly in the $(u,v,w)$
space.  The only difference is that the roles of $(v,y_v)$ and $(w,y_w)$
are swapped, and there is an extra factor multiplying the pure function,
corresponding to
\be
\frac{(1-y_v)(1-y_uy_vy_w)}{y_v (1-y_uy_w)} = \frac{\sqrt{\Delta}}{v} \,.
\label{Znewfactor}
\ee
Thus we get,
\bea
\tilde{V}_X + \tilde{f}
&=&  - \int_{1}^{u} \frac{du_t}{u_t(u_t-1)}
 \, \frac{\sqrt{\Delta(u_t,v_t,w_t)}}{v_t} 
 \, \tilde{Z}(u_t,v_t,w_t) \,,
\label{VtildenewA}\\
&=&  - \sqrt{\Delta(u,v,w)} \int_{1}^{u} 
\frac{du_t\ \tilde{Z}(u_t,v_t,w_t)}
 { v_t \bigl[ u \, (1-w) + (w-u) \, u_t \bigr] }  \,,
\label{VtildenewB}
\eea
where
\bea
v_t &=& 1 
- \frac{ (1-v) \, u_t \, (1-u_t) }{ u \, (1-w) + (w-u) \, u_t }  \,,
\label{vtdefZ}\\
w_t &=& \frac{(1-u) \, w \, u_t }{ u \, (1-w) + (w-u) \, u_t }
\,.
\label{wtdefZ}
\end{eqnarray}
The second form of the integral, \eqn{VtildenewB}, makes clear that
in the positive octant, $\tilde{V}_X+\tilde{f}$ is real for $\Delta > 0$,
and pure imaginary for $\Delta < 0$.  The overall sign of \eqn{VtildenewB}
corresponds to the branch of $\sqrt{\Delta}$ defined in term
of the $y$ variables in \eqn{rtdeltayrelation}; it ensures that the
logarithmic derivative with respect to $y_v$ reproduces $\tilde{Z}$.


\section{Ansatz and constraints at function level}
\label{sect-symboltofunction2}
\setcounter {equation} {0}

Now that we have obtained explicit functions representing the
symbols in section \ref{sect-ansatz}, we proceed to enumerate
the additional possible contributions, all of which have vanishing symbol.
The ratio function is real-valued in the Euclidean region in which
all three cross-ratios are positive.
Each of the above functions entering $V$ also has this property.
Therefore any additional functions that we add to our ansatz must also obey
the property.  In addition to the parameters
$\{ \alpha_X,\alpha_1,\ldots,\alpha_9 \}$,
we have the following real-valued parity-even beyond-the-symbol
ambiguities:
\begin{itemize}
\item At the $\zeta_{2}$ level,
\begin{align} \label{beyondsymbol1}
g^{(2)}=\zeta_{2} \, \bigl[&c_1 \, (\log^2 u + \log^2 w) + c_2 \log^2 v
+ c_3 \log(uw)\log v + c_4 \log u \log w \notag\\
&+ c_5 \, (H_2^u + H_2^w)  + c_6 H_2^v \bigr] \,, 
\end{align}
\item At the $\zeta_{3}$ level,
\begin{align}  \label{beyondsymbol2}
g^{(3)}= \zeta_{3} \, \bigl[ c_7 \log(uw) + c_8 \log v \bigr] \,,
\end{align}
\item At the $\zeta_{4}$ level,
\be \label{beyondsymbol3}
g^{(4)}=\zeta_{4} \, c_9\,.
\ee
\end{itemize}

If our ansatz is correct, then we expect that the parity-even function
$V$ should be given by
\be
V = \alpha_X \, V_X + \sum_{i=1}^9 \alpha_i \, f_i
+ g^{(2)} + g^{(3)} + g^{(4)} \,,
\label{completeVansatz}
\ee
for some rational values of the $\alpha_i$ and $c_i$.
There are no parity-odd beyond-the-symbol ambiguities that possess
only physical branch cuts.  (This fact follows from the absence of
an integrable parity-odd degree-two symbol whose first slot is constrained
to be $u$, $v$ or $w$.)

Next we would like to apply the constraint~(\ref{collinear})
from the collinear limit, namely $V(u,v,w) + V(w,u,v) \to 0$ as
$w\to 0$,\ $v\to 1-u$, but now at the level of functions, not just symbols.
One way to do this is to first complete the functions $V_X$ and $f_i$
into new functions $F_X = V_X + \hat{V}_X$, $F_i = f_i + \hat{f}_i$,
each of which gives a vanishing contribution to $V(u,v,w) + V(w,u,v)$
in the collinear limit. Although the symbols of the functions
$f_{1}, \ldots, f_{9}$ were already constrained to give a vanishing
contribution in this limit, that does not mean that they vanish as
functions.  Instead we will correct $V_X$ and the $f_{i}$ by appropriate
beyond-the-symbol terms, $\hat{V}_X$ and $\hat{f}_i$, which are
constructed from the expressions~(\ref{beyondsymbol1}), (\ref{beyondsymbol2})
and (\ref{beyondsymbol3}) for suitable values of the constants $c_i$.
The function $f_{9}$ requires no such correction, because it is the
two-loop MHV remainder function, which vanishes in all collinear limits.

To perform this correction, we need to know the collinear limits of
the functions $V_X$ and $f_i$.  For all but $f_8$, these limits are
straightforward to compute.  The limit of $f_{8}$ is more complicated to
obtain due to the presence of $\Omega^{(2)}$.  We compute this limit directly
in appendix~\ref{app-collinear}.  However, we may also observe
that in the collinear constraint equation~(\ref{collinear}),
only the combination $\Omega^{(2)}(w,u,1-u) + \Omega^{(2)}(1-u,w,u)$
is needed for small $w$.   (To see this, we use the symmetry of
$\Omega^{(2)}$ under exchange of its first two arguments.) 
This combination appears on the right-hand side of
\eqn{remaindernewrepresentation}, evaluated in the limit 
$w\to0$, $v\to1-u$, along with $\Omega^{(2)}(u,1-u,w)$ and the
simpler function $\mathcal{R}_{6,{\rm rat}}^{(2)}(u,1-u,w)$.
Now the left-hand side of this equation, the two-loop remainder
function, vanishes in the limit.  Also,
$\Omega^{(2)}(u,1-u,0) = \Psi^{(2)}(u,1-u) = 0$, where
$\Psi^{(2)}(u,v)$ is the two-loop pentaladder function~\cite{Drummond:2010cz}.
Hence the limit of the pair of $\Omega^{(2)}$ functions appearing in
$f_8(u,v,w)+f_8(w,u,v)$ reduces to evaluating
$\mathcal{R}_{6,{\rm rat}}^{(2)}(u,1-u,0)$, using \eqn{R62ratdef}.
The explicit formulas for all the required beyond-the-symbol
functions $\hat{V}_X$ and $\hat{f}_i$ are given in
appendix~\ref{app-collinear}.

The collinear constraint at the level of functions fixes 7 out of the 9
beyond-the-symbol terms, leaving only the following combinations
which have vanishing collinear contributions:
\begin{eqnarray}
\tilde{g}_{1} &=& \zeta_{2} \left[ \zeta_{2} + H_{2}^v
 - H_{2}^u - H_{2}^w \right] \,, \\
\tilde{g}_{2} &=& \zeta_{2} \left[ -\zeta_{2} + 2 H_{2}^v
 + \log(u w) \log v - \log u \log w \right] \,.
\end{eqnarray}
The function $V$ should therefore be given by
\be
V(u,v,w) = \alpha_X F_X
+ \sum_{i=1}^9 \alpha_i F_i
+ \tilde{c}_1 \tilde{g}_1 + \tilde{c}_2 \tilde{g}_2\,,
\label{Vfg}
\ee
where $\tilde{c}_1$ and $\tilde{c}_2$ are arbitrary constants.

We can analyse the constraints coming from the
spurious pole condition~(\ref{spurious}) in a similar way. 
The end result of this analysis is that one more beyond-the-symbol
ambiguity is fixed, leaving just one such function free.
In fact, this is the maximum number of beyond-the-symbol terms we can fix
with this analysis, because one can always add $\zeta_{2}$ multiplied by
the one-loop ratio function, $V^{(1)}$, given in \eqn{Voneloop}.
This product is a linear combination of $\tilde{g}_1$ and $\tilde{g}_2$,
namely $\tilde{g}_2 - \tilde{g}_1$,
and it automatically satisfies all constraints by itself.
Thus the only remaining beyond-the-symbol ambiguity is
$\zeta_{2} \, V^{(1)}$.

In the next section, we will calculate analytically the loop integrals
contributing to the two-loop NMHV amplitude for special kinematics.
We will use this information to determine the remaining unfixed
parameters, $\alpha_X$, $\alpha_1$ through $\alpha_9$, and
(one of) $\tilde{c}_1$ and $\tilde{c}_2$.


\section{Analytic calculation using loop integrals}
\label{sect-analytic}
\setcounter {equation} {0}

In this section, we will fix the remaining undetermined parameters in our
ansatz by computing the ratio function analytically in a certain
kinematical regime.

We find it convenient to perform our calculation using a mass
regulator~\cite{Alday:2009zm}.  As was reviewed in section~\ref{def-nmhv},
the ratio function is infrared finite.  Moreover, it should be independent
of the regularization scheme used to compute it.  We first verify
this statement at one loop by re-evaluating the MHV and NMHV six-point
amplitudes in the mass regularization.  At two loops, we find agreement
with previous numerical results~\cite{Kosower:2010yk} obtained using
dimensional regularization.


\subsection{Review of six-point MHV amplitudes}

Recall from section~\ref{def-nmhv} that supersymmetry allows to write any MHV 
amplitudes to all loop orders as a product of the tree-level amplitude,
multiplied by a helicity-independent function. We have
\be
\mathcal{A}_{\rm MHV}(a)\ =\ \mathcal{A}_{\rm MHV}^{(0)} \times M(a)\,,
\label{Mdef}
\ee
where $M(a) = 1 + a M^{(1)} + a^2 M^{(2)}+ \ldots$,
and $a$ is defined in \eqn{a_def}.
The known structure of infrared divergences takes a particularly simple
form if we consider $\log M$, namely~\cite{Henn:2010ir}
\begin{eqnarray}\label{all-loop-MHV1}
\log M(a,x_{ij}^2) &=& \sum_{i=1}^{6} \left[
- \frac{\gamma(a)}{16} \log^2  \frac{x_{i,i+2}^2}{m^2}
- \frac{\tilde{\mathcal G}_{0}(a)}{2} \log \frac{x_{i,i+2}^2}{m^2}
+ \tilde{f}(a) \right] \nonumber + F(a,x_{ij}^2) + \cO(m^2) \,.
\end{eqnarray}
Here $\gamma(a)$ is the cusp anomalous
dimension~\cite{Korchemskaya:1992je}. It is given by
\begin{eqnarray}
\gamma(a) &=& 4 a - 4 \zeta_2 a^2 + \cO(a^3) \,,
\end{eqnarray}
and we have
\begin{eqnarray}
\tilde{\mathcal G}_{0}(a) &=& -\zeta_{3} a^2 + \cO(a^3) \,, \\
\tilde{f}(a) &=& \frac{\zeta_{4}}{2} a^2 + \cO(a^3) \,.
\end{eqnarray}
Moreover, the finite part $F$ satisfies a dual conformal Ward
identity~\cite{Drummond:2007cf,Drummond:2007au}, whose most general
solution is
\begin{eqnarray}\label{all-loop-MHV2}
F(a,x_{ij}^2) &=& \frac{1}{4} \gamma(a) F^{(1)}(x_{ij}^2)
+ {\mathcal R}_6(u,v,w; a) + \tilde{C}(a) + \cO(m^2)\,,
\end{eqnarray}
with
\begin{eqnarray}
\tilde{C}(a) &=&-\frac{5 \zeta_{4}}{4} a^2 + \cO(a^3) \,.
\end{eqnarray}
The first term on the right-hand side of \eqn{all-loop-MHV2} 
comes from the BDS ansatz~\cite{Bern:2005iz}, and provides
a particular solution to the Ward identity. It is given by the one-loop
contribution to $F$, multiplied by one quarter of the
(coupling-dependent) cusp anomalous dimension $\gamma(a)$.
Hence its kinematical dependence is determined by the one-loop result.
The second term on the right-hand side of \eqn{all-loop-MHV2}, the remainder
function ${\mathcal R}_6(u,v,w;a)$~\cite{Bern:2008ap,Drummond:2008aq},
depends on three conformal cross-ratios, $u$, $v$ and $w$.
There is no remainder function for four and five points, because
non-vanishing conformal cross-ratios only appear starting at six points.
The specific choice of the kinematic-independent terms $\tilde{f}(a)$
and $\tilde{C}(a)$, determined by the four-point and five-point cases,
was made in such a way~\cite{Bern:2005iz} that ${\mathcal R}_6(u,v,w; a)$
vanishes in the collinear limit.

\subsection{Six-point NMHV amplitudes and ratio function}

The NMHV amplitude can be written as
\be \label{nmhv-all-loops}
\mathcal{A}_{\rm NMHV}(a)
= \frac{1}{2} \mathcal{A}^{(0)}_{\rm MHV}
\left[ [ (2) + (5) ] \, W_{1}(a)
     \ -\ [ (2) - (5) ] \, \tilde{W}_{1}(a)\ +\ {\rm cyclic} \right]\,,
\ee
where $W_{1}(a) = 1 + a W_1^{(1)}+ a^2 W_1^{(2)} + \ldots$ and
$\tilde{W}_1(a) = a^2 \tilde{W}_{1}^{(2)} + \ldots$.
Cyclic symmetry implies that under a cyclic rotation $\mathbb{P}$ of the
external legs, $i\to i+1$, the $W_i$ permute into each other according to
$\mathbb{P} W_{1} = W_{2}$, $\mathbb{P}^2 W_{1} = W_{3}$,
$\mathbb{P}^3 W_{1} = W_{1}$,
and similarly for the $\tilde{W}_{i}$.

We recall from section~\ref{def-nmhv}, \eqn{ratio123}, that
the ratio function(s) $V_{i}$ and $\tilde{V}_{i}$ are defined
by~\cite{Drummond:2008vq}\footnote{The original
definition~\cite{Drummond:2008vq} differs from one used
later~\cite{Kosower:2010yk} by a (coupling-dependent)
constant.  We use the latter definition~\cite{Kosower:2010yk}
because it makes the collinear behavior of the $V_{i}$ simpler.
Note that $V_{i}$ is called $C_{i}$ in ref.~\cite{Kosower:2010yk}.}
\be \label{def-ratio1}
\mathcal{A}_{\rm NMHV}(a)= \frac{1}{2} \mathcal{A}_{\rm MHV}(a) 
\left[ [ (2) + (5)] \, V_{1}(a)\ -\ [ (2)-(5) ] \, \tilde{V}_{1}(a)
\ +\ {\rm cyclic} \right]\,.
\ee
Based on the universality of infrared divergences, and in particular
the independence of infrared divergences on the helicity
configuration, the ratio function $\cP$ defined in \eqn{ratiodef} is
expected to be infrared finite, and independent of the regularization
scheme used to compute it.  More explicitly, comparing
eqs.~(\ref{Mdef}), (\ref{nmhv-all-loops}) and (\ref{def-ratio1}),
we see that
\be
W_{i}(a) = M(a) \, V_{i}(a)\,,\qquad
\tilde{W}_{i}(a) = M(a) \,\tilde{V}_{i}(a) \,, \qquad i=1,2,3 \,.
\ee
Expanding these relations in the coupling constant,
we find, at the one- and two-loop orders,
\begin{eqnarray}\label{definitionV1}
V^{(1)}_{i}  &=& W_{i}^{(1)} - M^{(1)}   \,, \\
V^{(2)}_{i} &=& W_{i}^{(2)} - M^{(2)} - M^{(1)} V^{(1)}_{i}   \,,
\label{definitionV2} \\
\tilde{V}^{(2)}_{i} &=& \tilde{W}_{i}^{(2)} \,.
\label{definitiontildeV2}
\end{eqnarray}
It will be a non-trivial check of our calculation that all infrared
divergences cancel in $V^{(2)}_{i}$.

\subsection{The one-loop ratio function}

At one loop, the MHV amplitude is given by~\cite{Bern:1994zx}
\be\label{MHV1loop}
M^{(1)} = -\frac{1}{8} \sum_{\sigma \in S_{1}
    \cup\, {\mathbb{P}} S_{1} \cup \, {\mathbb{P}}^2 S_{1}  }
\left[ F^{1m}(\sigma) -\frac{1}{2} F^{2me}(\sigma) \right] + \cO(m^2)\,,
\ee
where $S_1 = \left\{ (123456),(321654),(456123),(654321)\right\}$ and
$(abcdef)$ denotes a permutation of the external momenta.
In the NMHV case, we have~\cite{Bern:1994cg} 
\be\label{NMHV1loop}
W_1^{(1)} = - \frac{1}{4} \sum_{\sigma \in S_1 }
 \left[  F^{1m}(\sigma) +  F^{2mh}(\sigma) \right] + \cO(m^2) \,.
\ee
In writing~\eqns{MHV1loop}{NMHV1loop}, we converted the corresponding
expressions in dimensional regularization to mass regularization.  The
definitions of the integrals $F^{1m}, F^{2me}$ and $F^{2mh}$ in this
regularization are given in appendix~\ref{app-integrals}.

Inserting these results into \eqn{definitionV1} to obtain $V_1^{(1)}$,
and then applying the permutation ${\mathbb{P}}^2$ to get
$V_3^{(1)} \equiv V^{(1)}(u,v,w)$, we recover the expressions for
$V^{(1)}$ and $\tilde{V}^{(1)}$ in \eqns{Voneloop}{tildeVoneloop}.
These results are in perfect agreement with the results of an
earlier computation using dimensional regularization~\cite{Kosower:2010yk},
confirming the expectation that the ratio function 
should be independent of the regularization scheme.
(The result of the original calculation~\cite{Drummond:2008vq} of
the one-loop ratio function differs by a convention-dependent constant.)

Let us check that the collinear and spurious conditions reviewed in
section~\ref{def-nmhv} are satisfied.  They are given
by~\eqns{collinear}{spurious}, respectively. Indeed, we have that
\be
\lim_{w \to 0} \; \left[  V^{(1)}(u,1-u,w) + V^{(1)}(w,u,1-u) \right] = 0\,, 
\ee
and
\be
V^{(1)}(u,v,1) - V^{(1)}(1,u,v) = 0 \,.
\ee


\subsection{The two-loop ratio function}

There exist several representations of $M^{(2)}$ and $W_{1}^{(2)}$ in
terms of loop integrals.  Using generalized unitarity and dimensional
regularization, representations for the loop integrand of $M^{(2)}$ and
the even part of $W_{1}^{(2)}$ were found in
refs.~\cite{Bern:2008ap} and~\cite{Kosower:2010yk}, respectively. Alternative
expressions for a four-dimensional integrand were derived using on-shell
recursion relations in refs.~\cite{ArkaniHamed:2010kv,ArkaniHamed:2010gh}.
This loop integral representation also describes the odd part
$\tilde{W}_{1}^{(2)}$.  However, it will be convenient for us to
choose a form in which the MHV and NMHV amplitudes are treated in a
uniform way~\cite{Bern:2008ap,Kosower:2010yk}.

As in the one-loop case, we will assume that the loop integrals appearing
in the massive regularization are the analogs of those appearing in
dimensional regularization~\cite{Kosower:2010yk}.  A similar assumption
was made for the four-point amplitude up to four
loops~\cite{Henn:2010bk,Henn:2010ir},
and for the two-loop MHV amplitudes up to six
points~\cite{Drummond:2010mb}.  The latter work also required
promoting the planar four-dimensional loop integrands of 
ref.~\cite{ArkaniHamed:2010kv} into objects that can be integrated
to give a finite result.  We should point out that
this procedure could in principle miss terms whose integrand vanishes as
the mass vanishes, $m^2 \to 0$, but that are finite after integration.
Although examples of 
such integrals have been given~\cite{Henn:2010ir,Henn:2010kb}, they have
not yet proved relevant in a practical calculation.  In principle, there
are various ways of introducing mass regulators, which differ
in how masses are given to different propagators, leading to different
results after integration.  We will use the mass regulator of
ref.~\cite{Alday:2009zm}, which provides a systematic way of introducing
the masses.

We should also comment on `$\mu$-integrals' present in dimensional
regularization, in which numerator factors involve explicit factors of
the extra-dimensional components $\vec{\mu}$ of the loop momentum.
These integrals do not seem to have an analog in mass regularization,
at least when one neglects terms that vanish as $m^2 \to 0$.  The
$\mu$-integrals arise in dimensional regularization due to a mismatch
in dimension between the four-dimensional external polarization
vectors and the $D$-dimensional loop integration variable. It has been
observed in explicit computations that in the quantity $\log M$ the
$\mu$-integrals only contribute at $\cO(\epsilon)$ in dimensional
regularization. At two loops, this requires a cancellation involving
one- and two-loop $\mu$-integrals~\cite{Bern:2008ap}.  Such an
interference has no analog, at least through $\cO(m^2)$, in the
massive regularization, and therefore we drop the $\mu$-integrals in
the dimensionally-regularized integrands of refs.~\cite{Bern:2008ap}
and~\cite{Kosower:2010yk}.

\begin{figure}[t]
\centerline{\includegraphics[scale=0.8]{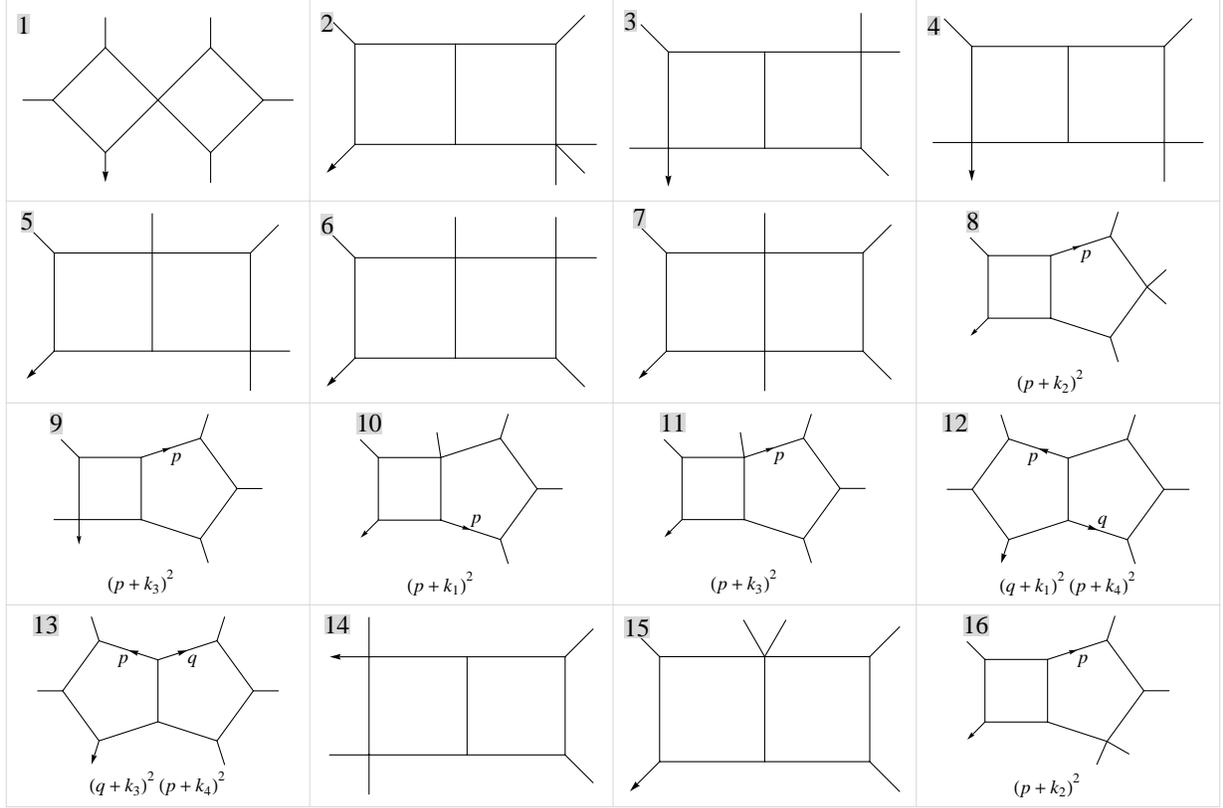}}
\caption{Two-loop integrals $I^{(i)}$ entering the two-loop six-point
MHV and NMHV amplitudes.  The labels $i$ are to the upper left of
each graph.  Solid internal lines indicate scalar propagators, while
numerator factors (if any) are shown below the graph.
The arrow on the external line indicates leg number $1$.
The figure is from ref.~\cite{Kosower:2010yk}.
\label{2loopints}}\nonumber
\end{figure}

At two loops, both the MHV amplitude and the even part of the NMHV
amplitude can be parametrized by~\cite{Kosower:2010yk}
\bea
S^{(2)}&=& \frac{1}{4}c_1 I^{(1)} + c_2 I^{(2)} + \frac{1}{2}c_3 I^{(3)}
+ \frac{1}{2} c_{4} I^{(4)} + c_{5} I^{(5)} + c_{6} I^{(6)}
\nonumber
\\[.5pt]
&+& \frac{1}{4} \left(
   c_{7a}    \mathbb{P}^{-2} I^{(7)} + c_{7b} \mathbb{P}^{-1} I^{(7)}
+ c_{7c}  I^{(7)}\right) + \frac{1}{2} c_{8} I^{(8)} + c_{9} I^{(9)}
\nonumber
\\[.5pt]
&+&
c_{10} I^{(10)} + c_{11} I^{(11)} + \frac{1}{2} c_{12} I^{(12)}
+ \frac{1}{2} c_{13} I^{(13)}
\nonumber
\\[.5pt]
&+& \frac{1}{2} c_{14} I^{(14)} + \frac{1}{2} c_{15} I^{(15)}
+ c_{16} I^{(16)}\,.
\label{twoloopS}
\eea
The integrals $I^{(i)}$ that enter are depicted in \fig{2loopints}.
We recall that $\mathbb{P}$ denotes a rotation of the external momenta by
one unit.  The coefficients $c_i$ are given by
\be
\begin{array}{rlcrl}
c_1 =&
s_{123} \bigl(s_{12} s_{45} s_{234}  + s_{23} s_{56} s_{345}
&~~~~~~&
c_2 =&
2 s_{23} s_{12}^2 \displaybreak[0] \\
&
+s_{123} (s_{34} s_{61} - s_{234} s_{345})\bigr)
&~~&          &   \displaybreak[0]\\
c_3 =&
s_{123} (s_{345} s_{123} - s_{45} s_{12})
&&
c_4 =&
s_{34} s_{123}^2 \displaybreak[0]\\
c_5 =&
s_{12} (s_{234} s_{123} - 2 s_{23} s_{56})
&&
c_6 =&
- s_{61} s_{12} s_{123} \displaybreak[0]\\
c_{7a} =&
 s_{123} (s_{234} s_{345}  - s_{34} s_{61})
&&
c_{7b} =&
 - 4 s_{34} s_{61} s_{123} 
\displaybreak[0]\\
c_{7c} =&
 s_{123}( s_{234} s_{345} - s_{34} s_{61})
&&
c_8 =&
2 s_{12} (s_{345} s_{123} - s_{12} s_{45}) \displaybreak[0]\\
c_9 =&
s_{45} s_{56} s_{123}
&&
c_{10} =&
s_{56} (2 s_{12} s_{45} - s_{123} s_{345})\displaybreak[0] \\
c_{11} =&
s_{61} s_{56} s_{123}
&&
c_{12} =&
s_{123} (s_{345} s_{123} - s_{12} s_{45})\displaybreak[0] \\
c_{13} =&
- s_{123}^2 s_{61}
&&
c_{14} =&
0 \displaybreak[0]\\
c_{15} =&
0
&&
c_{16} =&
0\end{array}
\label{IntegralCoefficientsMHV}
\ee
for the MHV case~\cite{Bern:2008ap}, and by
\be
\begin{array}{rlcrl}
c_1 =& - s_{1 2 3}^2 s_{3 4} s_{6 1} + s_{1 2 3}^2 s_{2 3 4} s_{3 4 5}  
&~~~& c_2 =& 2 s_{1 2}^2 s_{2 3} \cr
 & - s_{1 2 3} s_{2 3 4} s_{1 2} s_{4 5}  -s_{1 2 3} s_{3 4 5} s_{2 3} s_{5 6}
   + \! 2 s_{1 2} s_{2 3} s_{4 5} s_{5 6}
&&          &  \cr
c_3 =& s_{1 2 3} (s_{1 2 3} s_{3 4 5} - s_{1 2} s_{4 5})
&& c_4 = & s_{1 2 3}^2 s_{3 4}  \cr
c_5 =& - s_{1 2} s_{1 2 3} s_{2 3 4}
&&c_6 =& s_{6 1} s_{1 2} s_{1 2 3}\cr
c_{7a} =& -s_{1 2 3}  (s_{3 4 5} s_{2 3 4} -
 s_{6 1} s_{3 4})
&&
c_{7b} =& 2s_{1 2 3} s_{3 4} s_{6 1} \cr
c_{7c} =&
-s_{1 2 3} (s_{2 3 4} s_{3 4 5} -
 s_{6 1} s_{3 4})
&& c_8 =& 0        \cr
c_9 =& s_{1 2 3} s_{4 5} s_{5 6}          && 
c_{10} =& s_{5 6} s_{1 2 3} s_{3 4 5}   \cr
c_{11} =& -s_{5 6} s_{6 1} s_{1 2 3}
&&
c_{12} =& -s_{1 2 3}( s_{1 2 3} s_{3 4 5} \! - \! s_{1 2} s_{4 5})
\cr
c_{13} =& s_{1 2 3}^2 s_{6 1}
&&
c_{14} =& 2 s_{3 4}^2 s_{1 2 3}
\cr
c_{15} =& 0
&&
c_{16} =& 2 s_{1 2} s_{3 4} s_{1 2 3}
\end{array}
\label{IntegralCoefficientsNMHV}
\ee
for the NMHV case~\cite{Kosower:2010yk}.
Here $s_{i,i+1} = x_{i,i+2}$ and $s_{i,i+1,i+2}=x_{i,i+3}$, with all
indices understood to be defined modulo 6.

Then we can write
\bea \label{defloopM2}
M^{(2)} &=&  \frac{1}{16} 
\sum_{\sigma \in S_{1} \cup\, \mathbb{P} S_{1} \cup \, \mathbb{P}^2 S_{1} }
  S^{(2)}_{\rm MHV}  + \cO(m^2) \,, \\
\label{defloopW2}
W_{1}^{(2)} &=&  \frac{1}{8} \sum_{\sigma \in S_{1}  }  S^{(2)}_{\rm NMHV}
 + \cO(m^2)  \,.
\eea
In refs.~\cite{Bern:2008ap,Kosower:2010yk}, the dimensionally-regularized
version of the above formulas was used to study these amplitudes
numerically. In particular, the dual conformal invariance of
the remainder and ratio functions was tested.
The individual integrals are rather complicated, especially the ones of
double-pentagon type, and an analytic formula for them is not known yet.

Let us discuss several strategies that might be used to simplify the
calculation.

In ref.~\cite{Drummond:2010mb}, the calculation of $M^{(2)}$ in the
massive regularization was simplified by going from the above integral
basis to a more convenient one.  In particular, the complicated
double-pentagon integrals were replaced by other double-pentagon
integrals (plus simpler integrals) that are conceptually and
practically easier to evaluate.

Another possibility is to exploit the fact that the ratio function is dual
conformally invariant, although the individual integrals contributing to
it are not. This fact can be used to simplify the expression for the ratio
function, by taking limits that leave the cross-ratios invariant, but
simplify the individual integrals.  This technique turned out to be very
useful in computing the Wilson loops dual to MHV
amplitudes~\cite{DelDuca:2010zg}.

Here we use a trick that relies on the following observation. The
ratio between the coefficients $c_{12}$ and $c_{13}$ is exactly
the same in the MHV and NMHV case ---
see~\eqns{IntegralCoefficientsMHV}{IntegralCoefficientsNMHV}.
There is still a small mismatch in those terms when
comparing \eqns{defloopM2}{defloopW2}, due to the different permutation
sums.  However, this mismatch disappears if we choose a symmetrical
kinematical configuration.  We can choose, for example,
\be
K = \{ x_{i,i+2}^2=1 \,, x_{i,i+3}^2=1/\sqrt{u} \} \,,
\qquad i=1,2,\ldots,6,
\label{symK}
\ee
which corresponds to setting all three cross-ratios equal to $u$.  As we
will see, this kinematical subspace is more than sufficient to fix the
remaining ambiguities of the ansatz in the preceding section.

For equal cross-ratios, taking into account the prefactors and different
numbers of permutations in \eqns{defloopM2}{defloopW2}, we see
that the sum of $W_{1}^{(2)}$ and $\tfrac{2}{3} \, M^{(2)}$ not only cancels
the contributions from $I^{(12)}$ and $I^{(13)}$, but cancels or simplifies
several other coefficients as well.  We can write
\be
W_{1}^{(2)}[K] = S^{(2)}_{*} - \frac{2}{3} M^{(2)}[K]\,,
\ee
where $S^{(2)}_{*}$ is defined according to \eqn{twoloopS},
with the new coefficients
\begin{eqnarray}
c^{*}_{i} = \left\{ 1,2,\frac{1-u}{u^{3/2}},\frac{1}{u},-1,0,
                 -\frac{1}{\sqrt{u}},\frac{1}{u}-1,\frac{1}{\sqrt{u}},
               1,0,0,0,\frac{1}{\sqrt{u}},0,\frac{1}{\sqrt{u}} \right\} \,,
\end{eqnarray}
and where we have combined $c_{7} \equiv c_{7a} + c_{7b} + c_{7c}$,
because the corresponding integrals are equal at the symmetrical point
$(u,u,u)$.  Given the known analytical result for $M^{(2)}$,
we only need to evaluate the integrals $I^{(i)}$ for 
$i=1,2,3,4,5,7,8,9,10,14,16$ in order to obtain $W_{1}^{(2)}[K]$.
We could even further simplify the latter
integrals using a more convenient integral
basis~\cite{Drummond:2010mb,ArkaniHamed:2010kv,ArkaniHamed:2010gh},
but this turns out not to be necessary for the present purpose.

Taking into account \eqn{definitionV2}, we have
\be \label{V2trick}
V^{(2)}[K] = S^{(2)}_{*} -\frac{5}{3} M^{(2)}[K]
 - M^{(1)}[K]\ V^{(1)}[K] \,.
\ee
Let us collect the relevant formulas here, using in particular
\eqns{F1msym}{F2mesym}, and letting $L \equiv \log m^2$:
\begin{eqnarray}
V^{(1)}[K]  &=& \frac{1}{2} \log^2 u + \frac{3}{2} \, \Li_{2}(1-u)
 - \zeta_{2} \,, \\
M^{(1)}[K] &=&-\frac{3}{2} L^2 + \frac{\pi^2}{2}-\frac{3}{4} \log^2 u
 - \frac{3}{2} \, \Li_{2}\left(1-u \right)\,,\\
M^{(2)}[K]  &=& (\log M)^{(2)} + \frac{1}{2} (M^{(1)}[K] )^2 \nonumber \\
&=&   \frac{3}{2}\zeta_{2} L^2 -3 \zeta_{3} L + \frac{7}{4}\zeta_{4}
 -\zeta_{2} F^{(1)}[K] + {\mathcal R}_{6}^{(2)}(u,u,u)
 + \frac{1}{2} (M^{(1)}[K] )^2 \,,\\
F^{(1)}[K] &=& \frac{\pi^2}{2}-\frac{3}{4} \log^2 u
 - \frac{3}{2} \, \Li_{2}\left(1-u \right)\,.
\end{eqnarray}
We wish to emphasize that all terms appearing on the right-hand side of
\eqn{V2trick} have infrared divergences in the form of
powers of $L = \log(m^2)$, and that those terms must cancel in the
infrared-finite quantity $V^{(2)}$.  This cancellation is a
non-trivial check of our calculation.

The evaluation of the loop integrals proceeds in the standard way.  We
give a detailed example in appendix~\ref{appendix-twoloops}.  We derived
Mellin-Barnes representations for all integrals, and then used the
Mathematica code
{\sl MBasymptotics.m}~\cite{Czakon:2005rk,heptools1,heptools2}
in order to perform the asymptotic $m^2 \to 0$ limit. In this way we
could verify the cancellation of the infrared divergent terms,
analytically at the $L^4, L^3$ level, and numerically at the $L^2, L$ level.
The remaining finite $L^0$ terms are given by at most four-fold Mellin
Barnes integrals, which gives us a convenient way of evaluating
$V^{(2)}(u,u,u)$ numerically.

We can do better and use {\sl MBasymptotics.m} another time in order to
compute analytically the small $u$ and large $u$ limits of
$V^{(2)}(u,u,u)$. Having in mind that we want to fix the remaining
undetermined coefficients of our ansatz from section \ref{sect-ansatz}, we
go beyond the logarithmic terms in the expansion and also keep power
suppressed terms in $u$.

To promote these asymptotic limits back to a full function, we make the
analog of the ansatz of section~\ref{sect-ansatz}, by reducing the result
of section~\ref{sect-symboltofunction2} to the case of all cross-ratios
equal. Hence we expect $V^{(2)}(u,u,u)$ to be given by a linear
combination of $\mathcal{R}^{(2)}(u,u,u)$ and single-variable harmonic
polylogarithms.

We can then compare the asymptotic limits we computed against the
corresponding expansions of our ansatz.  In fact, when fitting a complete
function against just a few parameters it is highly non-trivial that we do
find a solution.  From comparing the first terms in the small $u$ and
large $u$ expansion (see appendix~\ref{appendix-twoloops} for more
details), we find
\begin{eqnarray}\label{eqVequalu}
V^{(2)}(u,u,u) &=&  -\frac{4}{3} {\mathcal R}_6^{(2)}
 + \frac{1}{16}\log^4 u 
 + \Bigl[ H_{2}^u - \frac{3}{2} \zeta_2 \Bigr] \log^2 u
- H_{3}^u \, \log u
+ \frac{1}{2} H_{4}^u + \frac{7}{4} H_{2,2}^u
 \nonumber \\
&&\null 
+ \frac{3}{2} \Bigl[ H_{2,1}^u \, \log u 
+ H_{3,1}^u + H_{2,1,1}^u - 3 \, \zeta_2 \, H_{2}^u \Bigr]
+ \frac{17}{3}\, \zeta_4 \,.
\end{eqnarray}
We also performed numerical checks of this expression at intermediate
values of $u$.  By comparing \eqn{eqVequalu} for 
$V^{(2)}(u,u,u) \equiv V(u,u,u)$ with our ansatz~(\ref{Vfg})
for $v=u$ and $w=u$, we find that the remaining twelve
parameters in the joint ansatz for $V(u,v,w)$ and 
$\tilde{V}^{(2)}(u,v,w)$ are all fixed.  (As mentioned in
section~\ref{sect-symboltofunction2}, there is one additional
constraint from the beyond-the-symbol spurious-pole constraint,
which is compatible with this solution.)  The values of the parameters are,
\be
\{ \alpha_X, \alpha_1, \ldots , \alpha_9 , \tilde{c}_1, \tilde{c}_2 \} 
\ =\ \{\tfrac{1}{8},
\tfrac{1}{4},\tfrac{3}{8},-\tfrac{5}{8},
-\tfrac{1}{4},\tfrac{1}{4},-\tfrac{1}{16},
0,\tfrac{1}{8},-1,
1,-1\}\,.
\label{alphacvalues}
\ee
We present the final form of the functions $V$ and $\tilde{V}$
in the next section.


\section{The final formula for the two-loop ratio function}
\label{sect-final}
\setcounter {equation} {0}

Now we insert the values of the twelve parameters that were fixed in the
previous section into our ansatz, and convert everything except
$\Omega^{(2)}$ and $\mathcal{R}_6^{(2)}$ into classical polylogarithms
whose arguments are simple, rational functions of $u$, $v$ and $w$.
The result is
\be
 V(u,v,w) = V^A(u,v,w) + V^A(w,v,u) + V^B(u,v,w),
\label{VABeq}
\ee
where
\bea
V^A(u,v,w) &=& 
- \frac{3}{4} \, \polylog_4\left(1-\frac{1}{u}\right)
- \polylog_4(1-u) + \ln u \, \polylog_3(1-u)
\nonumber\\ &&\null
- \frac{1}{4} \, \ln\left(\frac{uw}{v}\right)
      \, \biggl[ \polylog_3\left(1-\frac{1}{u}\right)
               + 2 \, \polylog_3(1-u) \biggr]
\nonumber\\ &&\null
+ \frac{1}{4} \, \polylog_2(1-v)
     \, \Bigl[ \polylog_2(1-u) + \ln u \, \ln v \Bigr]
\nonumber\\ &&\null
+ \frac{1}{8} \, \polylog_2(1-u) \, \Bigl[ 
  2 \, \polylog_2(1-u) - \ln^2 v - \ln^2 w 
   + 4 \, \ln v \, \ln w - 12 \, \zeta_2 \Bigr]
\,, \nonumber\\ 
&& ~ \label{VA}
\eea
and
\bea
V^B(u,v,w) &=& - \mathcal{R}_6^{(2)}(u,v,w)
- \frac{1}{4} \, \Omega^{(2)}(w,u,v)
\nonumber\\ &&\null
+ \frac{1}{8} \, \polylog_2(1-v) \Bigl[
     \polylog_2(1-v) - 2 \, \ln u \, \ln w - 8 \, \zeta_2 \Bigr]
\nonumber\\ &&\null
+ \frac{1}{4} \, \polylog_2(1-u) \, \polylog_2(1-w)
+ \frac{1}{16} \ln^2 v 
  \, \Bigl( \ln^2 u + \ln^2 w + 4 \, \ln u \, \ln w \Bigr)
\nonumber\\ &&\null
- \frac{1}{24} \, \ln v  \ln^3(uw)
+ \frac{1}{96} \ln^4(uw) - \frac{1}{16} \ln^2 u \, \ln^2 w
\nonumber\\ &&\null
+ \frac{\zeta_2}{4} \, \Bigl[ \ln^2 v 
 - 6 \, \Bigl( \ln v \, \ln(uw) - \ln u \, \ln w \Bigr) \Bigr]
+ 5 \, \zeta_4 \,.
\label{VB}
\eea

The function $\Omega^{(2)}$ can be evaluated as a simple one-dimensional
integral over classical polylogarithms with rational arguments,
using \eqns{Qphi1}{Omeganumnew} from section~\ref{sect-symboltofunction1}.
The function $\mathcal{R}_6^{(2)}$ is the two-loop remainder function.
It can be expressed entirely in terms of classical polylogarithms whose
arguments involve square-root functions of the cross
ratios~\cite{Goncharov:2010jf}.  Alternatively, it
can be expressed, using \eqn{remaindernewrepresentation},
in terms of three cyclic permutations of $\Omega^{(2)}$,
plus classical polylogarithms with rational arguments.
It is clear from \eqns{VA}{VB} that $V(u,v,w)$ is real in the positive octant,
given that $\mathcal{R}_6^{(2)}$ and $\Omega^{(2)}$ are.

For the odd part we find, using $\alpha_X = \alpha_{8} = \tfrac{1}{8}$,
\bea
\tilde{V}(u,v,w) =  \frac{1}{8} \, ( \tilde{V}_X +  \tilde{f} ) \,.
\label{Vtilde_answer}
\eea
This is exactly the linear combination of $\tilde{V}_X$ and
$\tilde{f}$ (multiplied by an overall $\tfrac{1}{8}$) for which we derived
a simple parametric integral formula in section
\ref{sect-symboltofunction1}.

\begin{figure}[t]
\centerline{\includegraphics[scale=0.8]{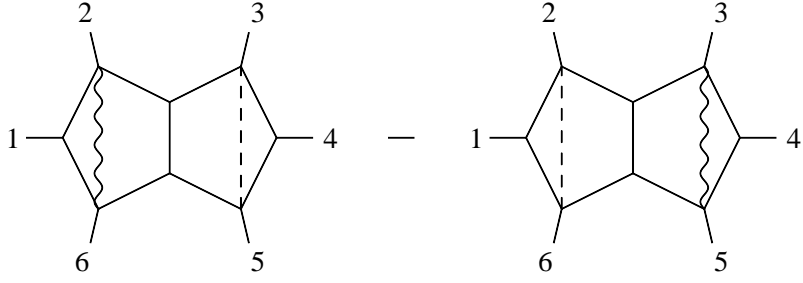}}
\caption{
Representation of $\tilde{V}$ in terms of the finite, dual conformal
loop integral $\tilde{\Omega}^{(2)}$.
This integral is evaluated in appendix~\ref{sect-computation-omega2tilde}.
The sum corresponds to the right-hand side of \eqn{newequationvtilde},
which also provides the proper overall normalization.}
\label{fig-vtilde}
\end{figure}

We can give an alternative form of the final answer that involves both
${\Omega}^{(2)}$ and the double-pentagon integral with `mixed'
numerator, $\tilde{\Omega}^{(2)}$ \cite{ArkaniHamed:2010kv}, where
the latter integral is evaluated in
appendix~\ref{sect-computation-omega2tilde}.  Due to the
respective symmetry and antisymmetry of $V$ and $\tilde{V}$ 
under exchange of their first and third arguments, \eqn{sym},
the NMHV ratio function is entirely specified by
$V(u,v,w) + \tilde{V}(y_u , y_v , y_w )$.  We have
\begin{align}\label{final-alternative}
V + \tilde{V} \;=\;  - \frac{1}{2} \, \Bigl[ \Omega^{(2)}(w,u,v) 
  + \tilde{\Omega}^{(2)}(1/y_w , 1/y_u , 1/y_v ) \Bigr]
 + T(u,v,w)  \,,
\end{align}
where $T(u,v,w)$ is implicitly defined by
eqs.~(\ref{remaindernewrepresentation}), (\ref{VABeq}) and
(\ref{result-omega2tilde}).  Explicitly it is given by,
\be
T(u,v,w) = T^A(u,v,w) + T^A(w,v,u) + T^B(u,v,w)\,,
\label{Tdef}
\ee
where
\bea
T^A(u,v,w) &=& - \frac{1}{2} \, \polylog_4\left(1-\frac{1}{u}\right)
  - \frac{3}{2} \, \polylog_4(1-u)
  + \frac{1}{2} \, \polylog_4(u) + \frac{1}{12} \, \ln^3 u \, \ln(1-u)
\nonumber\\ &&\hskip0cm \null
  + \ln\left(\frac{u v}{w}\right) \, \polylog_3(1-u)
  + \frac{1}{2} \, \ln\left(\frac{v}{w}\right)
                \, \polylog_3\left(1-\frac{1}{u}\right)
  + \frac{3}{8} \, \left[ \polylog_2(1-u) \right]^2
\nonumber\\ &&\hskip0cm \null
  + \frac{1}{8} \, \bigl[ 4 \, \polylog_2(1-u) + \ln^2 u \bigr]
         \, \polylog_2(1-v)
  + \frac{1}{8} \, \Bigl[ 6 \, \ln v \, \ln w
\nonumber\\ &&\hskip1cm \null
          - 2 \, \ln u \, \ln\left(\frac{v}{w}\right)
          - \ln^2 v - \ln^2 w - 12 \, \zeta_2 \Bigr] \, \polylog_2(1-u) \,,
\label{TAdef}
\eea
and
\bea
T^B(u,v,w) &=& \polylog_4\left(1-\frac{1}{v}\right)
  + \frac{1}{2} \, \polylog_4(1-v)
  + \frac{1}{2} \, \polylog_4(v) + \frac{1}{12} \, \ln^3 v \, \ln(1-v)
\nonumber\\ &&\hskip0cm \null
  + \frac{1}{2} \, \ln v \, \polylog_3\left(1-\frac{1}{v}\right)
  + \frac{1}{8} \, \left[ \polylog_2(1-v) \right]^2
  + \frac{1}{2} \, \polylog_2(1-u) \, \polylog_2(1-w)
\nonumber\\ &&\hskip0cm \null
  + \frac{1}{4} \, \bigl[ \ln(u w)\,\ln v - \ln u \, \ln w
              - 2 \, \zeta_2 \bigr]
             \, \bigl[ \polylog_2(1-v) - 6 \, \zeta_2 \bigr]
\nonumber\\ &&\hskip0cm \null
  - \frac{1}{48} \, \ln^4\left(\frac{u}{w}\right)
  + \frac{1}{16} \, \ln^2 u \, \ln^2 w
  - \frac{1}{12} \, ( \ln^3 u + \ln^3 w ) \, \ln v
\nonumber\\ &&\hskip0cm \null
  + \frac{1}{16} \, ( \ln^2 u + \ln^2 w + 4 \, \ln u \, \ln w ) \, \ln^2 v
  - \frac{1}{24} \, \ln^4 v
\nonumber\\ &&\hskip0cm \null
  - \frac{\zeta_2}{4} \, ( \ln^2 u + \ln^2 w - \ln^2 v )
  - \frac{\zeta_3}{2} \, \ln(u v w) - 3 \, \zeta_4 
\,.
\label{TBdef}
\eea
We see that $T$ is given by sums of products of
logarithms and polylogarithms with arguments which are rational
combinations of $u,v,w$.
In other words, the most complicated piece of $V + \tilde{V}$
is captured by the two double-pentagon integrals on the right-hand
side of equation (\ref{final-alternative}).

Moreover, the {\sl only} term containing parity-odd pieces on the
right-hand side of (\ref{final-alternative}) is
$\tilde{\Omega}^{(2)}$.  We can easily project out the parity-even
piece by taking a linear combination of this integral minus the same
integral rotated by three steps in the twistor variables.  This means
that we have an extremely simple representation of the parity-odd
function $\tilde{V}$ in terms of finite, dual conformal loop
integrals (see \fig{fig-vtilde}),
\begin{align}\label{newequationvtilde}
\tilde{V} = \frac{1}{4} \, \left[ 
\tilde{\Omega}^{(2)}( y_w , y_u , y_v ) 
- \tilde{\Omega}^{(2)}( 1/y_w , 1/y_u , 1/y_v ) \right] \,. 
\end{align}
The same double-pentagon integral with mixed numerator appears in the
representation of the NMHV loop integrand that was given in
Table 1 of ref.~\cite{ArkaniHamed:2010kv}. The latter integral
contains both an even and an odd part, although it is not
immediately obvious how to separate the two. For example, although the
penta-box integrals appearing in the representation of
ref.~\cite{ArkaniHamed:2010kv} of that amplitude contain odd
parts, it can be shown that the latter are only $\cO(m^2)$ when the
integrals are evaluated using a massive regulator~\cite{Alday:2009zm}; 
see ref.~\cite{Drummond:2010mb}.

We can perform a numerical check of our result for the (parity-even)
$\times$ (parity-even) part.  Using the values obtained for 
$\Omega^{(2)}$ in eqs.~(\ref{Omega2numKRV1}), (\ref{Omega2numKRV2}) and
(\ref{Omega2numKRV3}), we find that
\bea
\bigl[ V + {\mathcal R}_6^{(2)} \bigr]
(\tfrac{16}{5},\tfrac{112}{85},\tfrac{28}{17})
&=& 14.428955293631618492 \,, \\
\bigl[ V + {\mathcal R}_6^{(2)} \bigr]
(\tfrac{112}{85},\tfrac{28}{17},\tfrac{16}{5})
&=& 12.613874875030471932 \,, \\
\bigl[ V + {\cal R}_{6}^{(2)} \bigr]
(\tfrac{28}{17},\tfrac{16}{5},\tfrac{112}{85})
&=& 11.705797993389994692 \,,
\label{VnumKRVasympoints}
\eea
in agreement with the values given in Table I of
ref.~\cite{Kosower:2010yk}, to the numerical accuracy given there.
For reference, we also give the value of ${\mathcal R}_6^{(2)}$,
which is the same for all three points due to its symmetry,
\be
{\mathcal R}_6^{(2)}(\tfrac{16}{5},\tfrac{112}{85},\tfrac{28}{17})
\ =\ -3.655432869447587985 \,.
\label{R62numKRVpoint}
\ee

We also give the numerical values of the parity-odd function at
these three points.
Here we have to specify the $y$ values, or equivalently the branch
of the square root of $\Delta$ that we consider.  At the three
points, $\Delta$ is negative, $\Delta =  -1.1049134948096$.
We take the positive imaginary branch of the square root,
$\sqrt{\Delta} = 1.0511486549530 \, i$ in defining the $y$ values
through \eqn{ydef0}.
We then evaluate \eqns{VtildenewB}{Vtilde_answer} to obtain,
\bea
\tilde{V}(\tfrac{16}{5},\tfrac{112}{85},\tfrac{28}{17})
&=& \ \ \, 0.09053803091646201664 \, i \,, \\
\tilde{V}(\tfrac{112}{85},\tfrac{28}{17},\tfrac{16}{5})
&=& -0.12117656112226985895 \, i \,, \\
\tilde{V}(\tfrac{28}{17},\tfrac{16}{5},\tfrac{112}{85})
&=& \ \ \, 0.03063853020580784231 \, i \,.
\label{VtildenumKRVasympoints}
\eea
Note that these three values sum to zero.

In fact, although it is not apparent from the integral
form~(\ref{VtildenewB}), for general kinematics the function
$\tilde{V}$ obeys
\be
\tilde{V}(y_u,y_v,y_w) + \tilde{V}(y_v,y_w,y_u)
+ \tilde{V}(y_w,y_u,y_v) = 0\,.
\label{Vtilde_nototalantisym}
\ee
This relation is a consequence of our ansatz and the symmetry
condition~(\ref{sym}).  Given this symmetry condition, 
\eqn{Vtilde_nototalantisym} means that the totally antisymmetric
part of $\tilde{V}$ vanishes.  Even if there had existed functions
within our ansatz with a totally antisymmetric part, we could have
removed them simply by noting that they never contribute to the
ratio function~(\ref{PVform}), due to the
condition~(\ref{5bracketidentity}).

In an auxiliary plain text file accompanying this article, we provide
the degree-four symbols for the functions $V$, $\tilde{V}$,
$\Omega^{(2)}$, $\tilde\Omega^{(2)}$, $T$ and $Y$. In these files, a term
$a\otimes b\otimes c\otimes d$ is written as SB$(a,b,c,d)$.


\section{Conclusions and outlook}
\label{sect-conclusions}
\setcounter {equation} {0}

In this paper we have obtained the full analytic result for the
two-loop ratio function in planar $\mathcal{N}=4$ super Yang-Mills theory.
Our method assumed the existence of two pure functions, $V$ and $\tilde{V}$,
characterizing the ratio function, and was based on making an ansatz for
the letters entering their symbols.  We then further restricted
the ansatz by imposing physical constraints, such as the behaviour in
collinear and spurious regimes, and constraints coming from the
operator product expansion of Wilson loops, leaving only a small
number of undetermined parameters. The remaining parameters were fixed
by an analytic computation of the loop integrals that contribute to the
ratio function in particular kinematical regions.

We analysed the constraints in the collinear and spurious pole
limits. It is interesting that the spurious pole constraint involves both the
(parity-even) $\times$ (parity-even) and the (parity-odd) $\times$
(parity-odd) part of the ratio function. We found that, within our
ansatz, the (parity-odd) $\times$ (parity-odd) part is uniquely fixed
by the (parity-even) $\times$ (parity-even) part. In particular, it is
necessarily non-zero.

We were able to express the ratio function in terms of sums of
products of classical polylogarithms of rational arguments,
plus two relatively simple new functions.
The first is the parity-even double-pentagon integral $\Omega^{(2)}$.
The second is a new function $\tilde{V}$ describing the parity-odd sector,
but it is also related to the parity-odd part of a second double-pentagon
integral, $\tilde\Omega^{(2)}$.
Neither of these two additional functions can be
expressed in terms of classical polylogarithms; however, we have provided
simple parametric integral formulas for them, based on the differential
equations that the integrals obey.
We have checked our result for the (parity-even) $\times$
(parity-even) part of the ratio function by an analytic two-loop
computation (in a special kinematical regime) performed in the present
paper, as well as against numerical values in the literature.

Let us comment on the class of functions that can appear within our
ansatz.  We considered symbols that are built from the set of nine
letters $\{u,v,w,1-u,1-v,1-w,y_{u},y_{v},y_{w}\}$, with the physical
constraint that the first entry should be drawn from the set
$\{u,v,w\}$ only, to exclude non-physical branch cuts.  At degrees 1,
2, 3 and 4 there are 3, 9, 25 and 69 integrable parity-even symbols of
this kind.  At degree 3 and 4 there are also 1 and 6 parity-odd
integrable symbols. respectively.
As a byproduct of our analysis, we have a complete basis of functions
corresponding to the parity-even symbols through degree four, without
imposing any symmetries or collinear or spurious pole constraints.
Three of the degree-four functions are given by
$\Omega^{(2)}$ in its three orientations, while the remaining
functions are simple sums of products of single-variable harmonic
polylogarithms, such as $H_{0,1,0,1}(1-u)$.   The labels and the
argument are chosen such that only a physical branch cut starting from
$u=0$ is present. In general the labels can be any combination of
zeros and ones, provided that the last label is 1.
The unique parity-odd function at degree three is just the (rescaled)
six-dimensional hexagon integral $\tilde{\Phi}_{6}$, whose relevance
for scattering amplitudes in $\mathcal{N}=4$ super-Yang-Mills theory was
suggested earlier~\cite{Dixon:2011ng}.  The six parity-odd functions at
degree four are the three functions
$\tilde{\Phi}_{6} \log u$, $\tilde{\Phi}_{6} \log v$ and 
$\tilde{\Phi}_{6} \log w$; two more functions are given by $\tilde{V}$
in two orientations (which is also described by the parity-odd part of
the two-loop mixed hexagon $\tilde{\Omega}^{(2)}$); and there is one
further function.

Beyond two loops ({\it i.e.}~for symbols of degree higher than four)
new functions can appear, as in the three-loop MHV remainder
function~\cite{Dixon:2011pw}.  It would be very interesting to find
representations for them, analogous to the simple parametric integral
representations obtained in this paper.

The ansatz we made for the symbol was motivated by explicit results
for loop amplitudes \cite{Goncharov:2010jf,Dixon:2011pw} and loop
integrals \cite{Drummond:2010cz,Dixon:2011ng}.  Another motivation
comes from thinking in terms of twistor-space variables.  Our ansatz
implies that the letters of the symbol factorise into four-brackets of
momentum twistors.  This seems natural, because for six points (and hence
six twistors describing the scattering data) intersections of lines
and planes in twistor space always factorise into twistor
four-brackets.  At any rate, it would be very interesting if one could
prove or disprove our ansatz for the six-point remainder function and
ratio function at an arbitrary number of loops.
If the ansatz is valid to all loop
orders for six-point amplitudes in $\mathcal{N}=4$ super Yang-Mills theory,
then it is an extremely powerful constraint on the $S$ matrix of that theory.


\section*{Acknowledgments}
\setcounter {equation} {0}
We are grateful to Simon Caron-Huot for discussions about the symbol
of the ratio function.
The work of JMH was supported in part by the Department of Energy
grant DE-FG02-90ER40542, and that of LJD by the US Department of
Energy under contract DE--AC02--76SF00515.


\appendix

\section{Pure functions and symbols}
\label{app-symbols}
\setcounter {equation} {0}

We define a pure function of degree
(or weight) $k$ recursively, by demanding that its differential satisfies
\be
d\, f^{(k)} = \sum_{r} f_r^{(k-1)} d \log \phi_r\,.
\label{pure}
\ee
The sum over $r$ is finite and $\phi_r$ are algebraic functions.
This recursive definition is for all positive $k$; the only degree zero
pure functions are constants. The definition~(\ref{pure})
includes logarithms and classical polylogarithms, as well as other
iterated integrals, such as harmonic polylogarithms of
one~\cite{Remiddi:1999ew}
or more~\cite{Gehrmann:2000zt,Gehrmann:2001ck,Gehrmann:2001pz,Maitre:2005uu}
variables.

The {\sl symbol}~\cite{symbolsC,symbolsB,symbolsG}
${\mathcal S}(f)$ of a pure function
$f$ is defined recursively with respect to~\eqn{pure},
\be 
{\mathcal S}( f^{(k)} )
= \sum_r {\mathcal S}( f_r^{(k-1)} ) \otimes \phi_r\,.
\label{symbol}
\ee
If we continue this process until we reach degree 0, we find that 
${\mathcal S}( f^{(k)} )$ is an element of the
$k$-fold tensor product of the space of algebraic functions,
\be
{\mathcal S}( f^{(k)} ) =
\sum_{\vec \alpha} \phi_{\alpha_1} \otimes \ldots \otimes \phi_{\alpha_k}\,,
\label{symbolk}
\ee
where $\vec \alpha \equiv \{ \alpha_1,\ldots,\alpha_k \}$.
The symbol of a function loses information about which logarithmic
branch the function is on.  It also does not detect functions that are
transcendental constants multiplied by pure functions of lower degree;
such functions have zero symbol.  The symbol therefore
corresponds to an equivalence class of functions that differ in these
aspects.  Nevertheless, the symbol is extremely useful, because
complicated identities between transcendental functions defined by
iterated integrals become simple algebraic identities.

If a symbol can be expressed as a sum of terms, and all entries in each
term belong to a given set of variables, then we say that the
symbol can be factorised in terms of that set of variables.
In this paper we have assumed that the pure functions associated
with the NMHV six-point ratio function can be factorised in terms of the
set~(\ref{letters}). 
From the definition of the symbol, a term containing an entry which
is a product can be split into the sum of two terms, according to
\be
\ldots \otimes \phi_1 \phi_2 \otimes \ldots
= \ldots \otimes \phi_1 \otimes \ldots\
+\ \ldots \otimes \phi_2 \otimes \ldots \,.
\ee
Performing this factorisation is usually necessary to identify all
algebraic relations between terms.  It is often necessary to perform
the step again after taking a kinematic limit, because the algebraic
relations in the limit are different than for generic kinematics.

The elements of the symbol are not all independent, but are related
by the integrability condition $d^2 f^{(k)}=0$ for any function $f^{(k)}$.
The integrability relations can be described simply:
Pick two adjacent slots in the symbol
$\phi_{\alpha_i} \otimes \phi_{\alpha_{i+1}}$ and replace the
corresponding elements by the wedge product
$d \log \phi_{\alpha_i} \wedge d \log \phi_{\alpha_{i+1}}$ in every
term. The resulting expression must vanish.

The symbol also makes clear the locations of
the discontinuities of the function. If ${\mathcal S}( f^{(k)})$ is given
by \eqn{symbolk}, then the degree $k$ function $f^{(k)}$ has a
branch cut starting at $\phi_{\alpha_1}=0$.  The discontinuity across
this branch cut, denoted by $\Delta_{\phi_{\alpha_1}} f^{(k)}$,
is also a pure function, of degree $(k-1)$.  Its symbol is found
by clipping the first element off the symbol for $f^{(k)}$:
\be
{\mathcal S}( \Delta_{\phi_{\alpha_1}} f^{(k)} ) = \sum_{\vec \alpha}
\phi_{\alpha_2} \otimes \ldots \otimes \phi_{\alpha_k}\,.
\label{discsymbol}
\ee
It is instructive to check, for example, the vanishing of the double $v$
discontinuity for the $f_i$ functions in \eqn{doublevfi}, by inspecting
their symbols.  Using ${\cal S}(H_2^v) = -v \otimes (1-v)$ is enough
to show that $f_1$ through $f_4$ obey this relation.  Using
${\cal S}(2H_{2,1}^v + \log v \, H_2^v) = -v \otimes (1-v) \otimes v$
and ${\cal S}(H_{2,2}^v) = v \otimes (1-v) \otimes v \otimes (1-v) $ is
enough to establish it for $f_5$, and so on.

In general, taking discontinuities commutes with taking derivatives,
and both operations can be carried out at symbol level.  These facts make
it straightforward to verify, starting from \eqn{doublevVX}, that the
double $v$ discontinuity of $V_X/(2356)$ is annihilated by the
operator ${\cal D}$ defined in \eqns{D_defn}{Dpmuvw}.


\section{Details of the collinear limit}
\label{app-collinear}
\setcounter {equation} {0}

We give here beyond-the-symbol completions of the functions
$V_X,f_1,\ldots,f_8$ obeying the collinear limit constraint. We denote
the completed functions by $F_X = V_X + \hat{V}_X$ or $F_i = f_i+\hat{f}_i$, 
where the $f_i$ were given already in the main text. The 
collinearly-consistent completions of the functions $V_X$ and
$f_1,\ldots,f_7$ are simple to calculate. We find that we can choose
\begin{align}
\hat{V}_{X} \;=\; &  \frac{\zeta_2}{30} \Bigl[
15 \, ( \log^2 u + \log^2 w ) + 7 \, \log (u w) \log v
- 67 \, \log u \log w + 75 \, \log^2 v
\nonumber \\
&\hskip1cm
- 16 \, \Bigl( \text{Li}_2(1-u) + \text{Li}_2(1-w) \Bigr) \Bigr]
- 3 \, \zeta_3 \log ( u v w ) \,,
\displaybreak[0] \\
\hat{f}_{1} \;=\; & \frac{\zeta_2}{3} \Bigl[
\log(u w) \log v - \log u \log w 
- \text{Li}_2(1-u) - \text{Li}_2(1-w) \Bigr] \,,
\displaybreak[0] \\
\hat{f}_{2} \;=\;& \frac{\zeta_2}{2} \Bigl[ \log^2 u + \log^2 w
+ 4 \log u \log w + \log^2 v \Bigr] + \zeta_3 \log(u v w) \,,
\displaybreak[0] \\
\hat{f}_{3} \;=\;& \zeta_2 \Bigl[ \log(u w) \log v - \log u \log w \Bigr] \,,
\displaybreak[0] \\
\hat{f}_{4} \;=\;& \zeta_2 \Bigl[ \log^2(u w) + \log^2 v \Bigr] \,,
\displaybreak[0] \\
\hat{f}_{5} \;=\;& \frac{\zeta_2}{15} \Bigl[ 2 \log(u w) \log v
 - 2 \log u \log w  
- 11 \Bigl( \text{Li}_2(1-u) +  \text{Li}_2(1-w) \Bigr) \Bigr] \,,
\displaybreak[0] \\
\hat{f}_{6} \;=\;& \zeta_2 \Bigl[ \log(u w) \log v -  \log u \log w
+ 2 \Bigl( \text{Li}_2(1-u) + \text{Li}_2(1-w) \Bigr) \Bigr]
+ 2 \zeta_3 \log(u v w) \,, \displaybreak[0] \\
\hat{f}_{7} \;=\;& \frac{2}{5} \zeta_2 
\Bigl[ 4 \log(u w) \log v - 9 \log u \log w
+ 8 \Bigl( \text{Li}_2(1-u) + \text{Li}_2(1-w) \Bigr) \Bigr]
  + \zeta_3 \log(u v w) \,.
\end{align}
To define $\hat{f}_8$, the limit $w \to 0$ of $\Omega^{(2)}(w,u,1-u)$
is required. Analyzing the symbol of $\Omega^{(2)}(w,u,1-u)$,
one expects the following behavior as $w \to 0$,
\be
\lim_{w \to 0} \Omega^{(2)}(w,u,1-u) = 
\log^2 w \, q_{2}(u) + \log w \,  q_{3}(u) + q_{4}(u) + \cO(w)\,.
\ee
From the symbol of $\Omega^{(2)}$ we can determine the symbol of
the $q_{i}(u)$.  Therefore, the only ambiguities to be fixed are
beyond-the-symbol terms in the $q_{i}$, for which we can make an ansatz.
Then, we fix the latter by comparing against the asymptotic $w \to 0$ limit
of a Mellin-Barnes representation of $\Omega^{(2)}$. We find,
\begin{align}
q_{2}(u) &= \frac{1}{4} \log^2 u + \frac{1}{2} \text{Li}_2(1-u)
\,,\displaybreak[0]\\
q_{3}(u) &= - \text{Li}_2(1-u) \Bigl( \log u + \log(1-u) \Bigr)
- \log^2 u \log(1-u) + \zeta_2 \, \log u 
\nonumber \\
&\hskip0.6cm 
+ \text{Li}_3(1-u)-\text{Li}_3(u) + \zeta_3
\,,\displaybreak[0]\\
q_{4}(u) &= \frac{1}{2} \, \log^3 u \, \log(1-u)
+ \frac{3}{4} \, \log^2 u \, \log^2(1-u)
\nonumber \\
&\hskip0.6cm 
+ \frac{1}{2} \Big[ \log^2 u + 4 \, \log u \, \log(1-u) + 2 \, \zeta_2 \Bigr]
  \text{Li}_2(1-u)
+ \frac{1}{2} \, [\text{Li}_2(1-u)]^2
\nonumber \\
&\hskip0.6cm 
 + \text{Li}_3(1-u) \Bigl( \log(1-u) - \log u \Bigr)
+ \log u \, \text{Li}_3(u) 
- 3 \, \text{Li}_4(1-u) - \text{Li}_4(u)
\nonumber \\
&\hskip0.6cm 
- 3 \, S_{2,2}(u)
+ 3 \, \zeta_3 \log u  + \frac{7}{4} \, \zeta_4 \,.
\end{align}
Here $S_{2,2}(u) = H_{0,0,1,1}(u)$ is the Nielsen polylogarithm.

The other limit that is needed in \eqn{collinear} can be obtained by
the symmetry of $\Omega^{(2)}$ in the first two entries,
\bea
&&\Omega^{(2)}(1-u,w,u) = \Omega^{(2)}(w,1-u,u) = 
\log^2 w \, q_{2}(1-u) + \log w \,  q_{3}(1-u) + q_{4}(1-u) \,.
\nonumber\\
&~&
\eea
Using these limits, we can determine a correction to $f_{8}$ such that
$f_{8} + \hat{f}_{8}$ satisfies~\eqn{collinear},
\begin{eqnarray}
\hat{f}_{8} &=& \frac{\zeta_2}{3} \Bigl[ \log(u w) \log v
- \log u \log w - \text{Li}_2(1-u) - \text{Li}_2(1-w) \Bigr]
+ \zeta_3 \log(u v w) \,.~~~~~~
\end{eqnarray}
We found the following identity helpful,
\begin{eqnarray}
0 &=& S_{2,2}(u) + S_{2,2}(1-u)
 + \log(1-u) \text{Li}_3(u) +  \log u \, \text{Li}_3(1-u) 
 +  \frac{1}{4} \log^2 u \log^2(1-u)
\nonumber \\ && \null
 - \zeta_2 \log u \log(1-u) 
- \zeta_3 \, \bigl( \log u +  \log(1-u) \bigr) - \frac{\zeta_4}{4}\,.
\end{eqnarray}
We also have
\begin{eqnarray}
\hat{f}_{9} &=& 0 \,,
\end{eqnarray}
because $f_9 = \mathcal{R}_6^{(2)}$ vanishes in all collinear limits.


\section{One-loop integrals in massive regularization}
\label{app-integrals}
\setcounter {equation} {0}

All integrals in our paper are given in the mostly-plus metric,
so that the distances $x_{ij}^2$ are positive in the Euclidean region.

\begin{figure}[t]
\centerline{\includegraphics[scale=0.8]{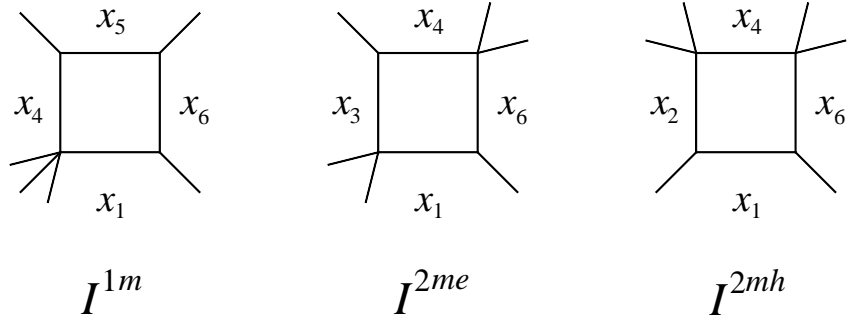}}
\caption{\small
One-loop box integrals appearing in MHV and NMHV amplitudes.}
\label{fig-one-loop-boxes}
\end{figure}

The integrals appearing in the one-loop MHV and NMHV amplitudes are
\begin{eqnarray}
I^{1m} &=&  \int \frac{d^{4}x_{j}}{i \pi^2}
\frac{1}{(x_{1j}^2 +m^2)  (x_{4j}^2+m^2) (x_{5j}^2+m^2) (x_{6j}^2+m^2)} \,, \\
I^{2me} &=& \int \frac{d^{4}x_{j}}{i \pi^2}
\frac{1}{(x_{6j}^2 +m^2)  (x_{1j}^2+m^2) (x_{3j}^2+m^2) (x_{4j}^2+m^2)} \,, \\
I^{2mh} &=& \int \frac{d^{4}x_{j}}{i \pi^2}
\frac{1}{(x_{6j}^2 +m^2)  (x_{1j}^2+m^2) (x_{2j}^2+m^2) (x_{4j}^2+m^2)} \,,
\end{eqnarray}
where we recall that $x_{i,i+1}^2=0$ with indices defined modulo $6$,
such that in particular $x_{61}^2=0$.  See \fig{fig-one-loop-boxes}.
It is convenient to define the dimensionless functions,
\begin{eqnarray}
F^{1m}&=&x_{46}^2 x_{15}^2 I^{1m} \,,\\
F^{2me} &=& (x_{13}^2 x_{46}^2-x_{14}^2 x_{36}^2) I^{2me}\,,\\
F^{2mh} &=& x_{14}^2 x_{26}^2 I^{2mh}\,.
\end{eqnarray}
They are given by
\begin{eqnarray}
F^{1m}&=& \log^2 \frac{m^2}{x_{46}^2} + \log^2 \frac{m^2}{x_{15}^2}
 - \log^2 \frac{m^2}{x_{14}^2} - \log^2 \frac{x_{15}^2}{x_{46}^2}
 - \frac{\pi^2}{3} \nonumber \\
&& - 2 \, \Li_2\left(1- \frac{x_{14}^2}{x_{15}^2}\right)
 - 2 \, \Li_2 \left( 1- \frac{ x_{14}^2 } {x_{46}^2} \right) + \cO(m^2)\,,
\end{eqnarray}
\begin{eqnarray}
F^{2me} &=& - \log^2 \frac{m^2}{x_{14}^2} - \log^2 \frac{m^2}{x_{36}^2}
 + \log^2 \frac{m^2}{x_{13}^2} + \log^2 \frac{m^2}{x_{46}^2}
 + \log^2 \frac{x_{14}^2}{x_{36}^2} \nonumber \\
&& + 2 \, \Li_2\left(1-\frac{x_{13}^2}{x_{14}^2}\right)
 + 2 \, \Li_2\left(1-\frac{x_{13}^2}{x_{36}^2}\right)
 + 2 \, \Li_2\left(1-\frac{x_{46}^2}{x_{14}^2}\right) \nonumber \\
&& + 2 \, \Li_2\left(1-\frac{x_{46}^2}{x_{36}^2}\right)
   - 2 \, \Li_2\left(1-\frac{x_{13}^2 x_{46}^2}{x_{14}^2 x_{36}^2}\right)
 + \cO(m^2)\,,
\end{eqnarray}
and
\begin{eqnarray}
F^{2mh} &=& \frac{1}{2}
\log^2\left( \frac{m^2 x_{24}^2 x_{46}^2}{(x_{14}^2)^2 x_{26}^2} \right)
- \log^2 \frac{x_{24}^2}{x_{14}^2} - \log \frac{x_{46}^2}{x_{14}^2}
 \nonumber \\
&& - 2 \, \Li_{2}\left(1-\frac{x_{24}^2}{x_{14}^2} \right)
   - 2\, \Li_{2}\left(1-\frac{x_{46}^2}{x_{14}^2} \right) + \cO(m^2) \,.
\end{eqnarray}
In the symmetric kinematics~(\ref{symK}), and neglecting the
$\cO(m^2)$ terms, we have,
\bea
F^{1m}&=& L^2 - 4 \, {\rm Li}_2\left(1-\frac{1}{\sqrt{u}}\right)
  - 2 \, \zeta_2 \,, \label{F1msym} \\
F^{2me}&=& 8 \, {\rm Li}_2(1-\sqrt{u}) - 2 \, {\rm Li}_2(1-u) \,,
\label{F2mesym} \\
F^{2mh}&=& \tfrac{1}{2} L^2 + L \, \log u - 4 \, {\rm Li}_2(1-\sqrt{u}) \,.
\label{F2mhsym}
\eea
%


\section{Description of the two-loop computation}
\label{appendix-twoloops}
\setcounter {equation} {0}

Let us illustrate the analytic computation of the loop integrals by
using the pentabox integral $I^{(10)}$ of \fig{2loopints}.
It is defined by
\begin{eqnarray}\label{int10example}
I^{(10)} &=&
\int \frac{d^{4}x_{i} d^{4}x_{j}}{ (i \pi^2)^2}
\frac{(x_{1j}^2 + m^2)}{(x_{6i}^2 + m^2) (x_{1i}^2 + m^2) 
              (x_{2i}^2 + m^2) x_{ij}^2}  \nonumber \\
&& \hspace{2cm} \times \frac{1}{(x_{2j}^2 + m^2)(x_{3j}^2 + m^2)
              (x_{4j}^2 + m^2)(x_{5j}^2 + m^2)}\,,
\end{eqnarray}
where the external dual coordinates are in the order $x_2,x_3,x_4,x_5,x_6,x_1$,
reading counter-clockwise from the external momentum $k_1$ in the figure.
Also, $x_i$ is the dual coordinate for the box, and $x_j$ is the one for the
pentagon. We remind the reader that the specific mass assignment in
\eqn{int10example}, in particular the fact that the internal propagator is
massless, follows from extended dual conformal symmetry.
See refs.~\cite{Alday:2009zm} and~\cite{Henn:2010ir} for further
explanation.

We proceed by deriving a Mellin-Barnes (MB) representation for this
integral. This is done by first introducing Feynman parameters in order to
carry out the four-dimensional loop integrations. Subsequently, MB
parameters are introduced to factorize the Feynman denominator, after
which the Feynman integrals can be done trivially. Experience shows that
it is convenient to introduce the MB parameters loop by
loop~\cite{smirnov2006feynman}.  Very detailed derivations of MB
representations for integrals like $I^{(10)}$ in \eqn{int10example}
can be found in appendix A of ref.~\cite{Henn:2010ir}.

In the case of integral $I^{(10)}$, the numerator factor $(x_{1j}^2 + m^2)$
deserves a comment. We choose to treat the latter as an inverse
propagator.  In doing so, some of the formulas we need to use, such as
the Feynman parameter formula, develop spurious divergences. In order
to be able to still use these formulas, we work with the analytically
continued integral $I^{(10)}(\delta)$, where the numerator factor is
replaced by $(x_{1j}^2 + m^2)^{1-2 \delta}$ and the integration measure
is changed to $d^{4+2 \delta}x_{j}$. We will do our computation
for $\delta \neq 0$, where all manipulations are allowed, and take the
$\delta \to 0$ limit later.  The MB representation we find in this way is
\begin{eqnarray}
I^{(10)}(\delta) &=&  (m^2)^{-3 - \delta}  \int \frac{dz_{i}}{(2 \pi i)^{12}} 
    \Gamma(-z_1 )  \left( \prod_{j=3}^{12} \Gamma(-z_{j})  \right)  
    \Gamma( 1 + z_1 ) \Gamma( 1 + z_1 + z_2 ) \nonumber \\
    &&\times  \Gamma(z_1 - z_2 - z_3 )  \Gamma(1 + z_3 )
   \Gamma( 1 + z_2 + z_3 )
   \Gamma( 2 + z_{10} + z_{11} + z_{12} + z_2 + z_3 ) 
     \nonumber \\
    &&\times
 \Gamma( z_{12} - z_2 + z_5 + z_6 )
 \Gamma( 1 + z_4 + z_5 + z_7 ) \Gamma( 1 + z_{10} + z_6 + z_8 ) 
    \nonumber \\
    &&\times
   \Gamma( 1 + z_{11} + z_4 + z_9 )
   \Gamma(-1 + 2 \delta - z_3 + z_7 + z_8 + z_9 )
   \Gamma( 2 + \delta + z_{4,12} )  \nonumber \\
        &&\times
    1/\bigl[ \Gamma( 2 + 2 z_{10} ) \Gamma(-1 + 2 \delta - z_3 )
           \Gamma( 2 + z_2 + z_3 ) 
           \Gamma\bigl( 2 (2 + \delta + z_{4,12}) \bigr) \bigr]
  \nonumber \\
    &&\times   
    \left(\frac{x^2_{13}}{m^2}\right)^{z_7} 
    \left(\frac{x^2_{14} }{m^2}\right)^{z_8} 
    \left(\frac{x^2_{15}}{m^2} \right)^{z_9} 
    \left(\frac{x^2_{2 4}}{m^2} \right)^{z_{10}}
    \left(\frac{x^2_{2 5}}{m^2} \right)^
   {z_{11}} 
   \left(\frac{x^2_{2 6}}{m^2} \right)^{z_1 + z_{12}} 
     \nonumber \\
    &&\times
   \left(\frac{x^2_{3 5}}{m^2} \right)^{z_4} 
   \left(\frac{x^2_{3 6}}{m^2} \right)^{z_{5}} 
   \left(\frac{x^2_{4 6}}{m^2} \right)^
   {z_6}  \,,
\end{eqnarray}
where $z_{4,12} = \sum_{j=4}^{12} z_j$.
Here the integrations go from $-i \infty$ to $i \infty$ in the complex
plane. The real part of the $z_{i}$ must be chosen such that the arguments
of all $\Gamma$ functions have positive real part. One finds that this is
only possible for $\delta \neq 0$.

The limit $\delta \to 0$ is very similar to the regulator limit in
dimensional regularization, with the difference that here we expect a
finite result, because the original integral was well-defined for
$\delta = 0$. In order to take the limit, one first has to deform some of the
$z_{i}$ integration contours~\cite{smirnov2006feynman}.
This procedure has been implemented in the {\sl MB.m} Mathematica
code~\cite{Czakon:2005rk,heptools1,heptools2}.

Having removed the auxiliary parameter $\delta$, we have a valid MB
representation for $I^{(10)}$.  We can now perform the regulator limit $m^2
\to 0$. This again involves deforming the integration contours, such that
the real part of the exponent of $m^2$ becomes positive, at which point a
Taylor expansion in $m^2$ is possible. We neglect power-suppressed terms
in $m^2$, since we are only interested in the logarithmic infrared
divergences and in the finite part.  In deforming the contours, one picks
up residues from poles of the $\Gamma$ functions, which can produce powers
of $\log m^2$. The resulting lower-dimensional integrals are treated in
the same way.

In fact, the leading divergent $\log^4 m^2 $ and $\log^3 m^2 $ terms are
obtained in this way without any remaining MB integrations. For example,
\bea
I^{(10)} &=& \frac{5}{8} \frac{1}{x_{24}^2 x_{26}^2 x_{35}^2} \log^4 m^2 
+\cO( \log^3 m^2) \,.
\eea
All $\log^{i} m^2$ terms with $i>0$ eventually cancel in the definition of
the remainder function. We will therefore focus on the finite terms as
$m^2 \to 0$. The latter are obtained as at most four-fold MB integrals.

In the main text, we have considered the special kinematical regime $K$
in \eqn{symK}, in which all three cross-ratios are equal to $u$.
It is easy to use the Mathematica
codes~\cite{Czakon:2005rk,heptools1,heptools2} in order to compute
the $u \to 0$ or $u \to \infty$ limits of $I^{(10)}[K]$ analytically.
For example, we find, in the small $u$ limit,
\bea \label{I10smalluexpansion}
\lim_{u \to 0} I^{(10)}[K]|_{\log^0 m^2} &=& \frac{3}{32} \log^4 u \nonumber \\
&& + \log^3 u \left[ \frac{5}{12}u^{1/2}+u+\frac{5 }{36}u^{3/2}
                +\frac{3 }{2}u^2+\frac{1}{12}u^{5/2}+\frac{10}{3} u^3
              + \cO(u^{7/2}) \right]  \nonumber \\
&& + \, \cO(\log^2 u )\,.
\eea 
It is straightforward to obtain higher orders in these expansions,
either analytically or numerically to high precision, but we refrain from
reproducing them here to save space.

Computing the asymptotic expansions of all integrals contributing to
$S^{(2)}_{*}$ in this way, we obtain
\begin{eqnarray}\label{Ssmallu}
\lim_{u \to 0} S^{(2)}_{*}|_{\log^0 m^2} &=& \frac{5}{32} \log^4 u
 \nonumber \\
&& + \, \log^3 u \left[ \frac{3}{4} u + \frac{7}{8} u^2 + \frac{7}{4} u^3
 + \frac{71}{16}  u^4 + \frac{253}{20} u^5+ \cO(u^6) \right]   \nonumber \\
&& + \, \log^2 u \left[ - \frac{\pi^2}{12}  + \frac{7}{4} u^2
  + \frac{19}{4} u^3 + \frac{653}{48} u^4 + \frac{995}{24} u^5
  + \cO(u^6) \right]   \nonumber \\
&& + \, \cO(\log u) \,,
\end{eqnarray}
in the small $u$ limit, and
\begin{eqnarray}\label{Sbigu}
\lim_{u \to \infty} S^{(2)}_{*}|_{\log^0 m^2} &=&
 \frac{1}{32} \log^4 u \nonumber  \\
&& \hspace{-2cm} - \, \log^3 u 
\left[ \frac{1}{24} u^{-1} + \frac{1}{48} u^{-2} + \frac{1}{72} u^{-3}
     + \frac{1}{96} u^{-4} + \frac{1}{120} u^{-5} + \cO(u^{-6}) \right]
\nonumber \\
&& \hspace{-2cm} + \, \log^2 u \left[  \frac{\pi^2}{24}
 + \frac{1}{16} u^{-1} - \frac{1}{64} u^{-2}- \frac{ 17}{1440} u^{-3}
 - \frac{67}{8960} u^{-4} - \frac{83}{16800} u^{-5} + \cO(u^{-6}) \right]
 \nonumber \\
&& \hspace{-2cm} + \, \cO(\log u)  \,,
\end{eqnarray}
in the large $u$ limit.  We remark that the half-integer powers appearing
in \eqn{I10smalluexpansion} have cancelled in the sum over all
integrals contributing to $S^{(2)}_{*}$.  Higher-order terms in the
expansions can be obtained numerically to great accuracy, but are not
displayed for brevity.

Comparing \eqns{Ssmallu}{Sbigu} to \eqn{V2trick}, we can fix
$V(u,u,u)+ \frac{5}{3} \,\mathcal{R}_{6}^{(2)}(u,u,u)$,
or equivalently $V(u,u,u)$, within our ansatz.
In this way we arrive at \eqn{eqVequalu} in the main text.

We can further test \eqn{eqVequalu} by using our four-fold MB
representation for $V(u,u,u)$ in order to compute some numerical values
at intermediate values of $u$. For example, we find
\begin{eqnarray}
V(\tfrac{1}{4},\tfrac{1}{4},\tfrac{1}{4})
&=& -3.49796 \pm 10^{-4}\,, \label{numcheck1}\\
V(12,12,12) &=& 35.56433 \pm 10^{-5}\,, \label{numcheck2}
\end{eqnarray}
using our MB representation of $V(u,u,u)$, and
\begin{eqnarray}
V(\tfrac{1}{4},\tfrac{1}{4},\tfrac{1}{4})
&=& -3.497905588766739\,,  \label{numcheck1b} \\
V(12,12,12) &=&  35.564326922499499\,, \label{numcheck1c}
\end{eqnarray}
using \eqn{eqVequalu}.

We also note that \eqns{numcheck1}{numcheck1b} agree, within the error
bounds, with the numerical value given in ref.~\cite{Kosower:2010yk}, namely
$V_{KRV}(\tfrac{1}{4},\tfrac{1}{4},\tfrac{1}{4}) = -3.502 \pm 0.002$.


\section{Computation of the mixed numerator integral $\tilde{\Omega}^{(2)}$}
\label{sect-computation-omega2tilde}
\setcounter {equation} {0}

\subsection*{Differential equation for $\tilde{\Omega}^{(2)}$}

We consider the double-pentagon integral with {\it mixed}
numerator~\cite{ArkaniHamed:2010kv},
\begin{align}\label{defomegatilde}
\tilde{\Omega}^{(2)}(y_u , y_v , y_w ) \; =&  \;
\int \frac{ d^{4}Z_{AB} d^{4}Z_{CD} }{(i \pi^2)^2}
 \frac{ (4612)(2346) (AB13)}{(AB61)(AB12)(AB23)(AB34)} \nonumber \\
& \; \qquad \qquad \qquad  
\times \frac{(CD (561) \cap (345))}{(ABCD)(CD34)(CD45)(CD56)(CD61)} \,,
\end{align}
where $(CD (561) \cap (345)) = (C561)(D345)-(D561)(C345)$.

Loop integrals of this type satisfy simple second-order differential
equations~\cite{Drummond:2010cz}.  The key point is the presence of
pentagon subintegrals that are also present in $\tilde{\Omega}^{(2)}$.
Following ref.~\cite{Drummond:2010cz}, it is easy to see that the latter
integral satisfies the differential equation
\begin{align}\label{diffeqomega2tilde}
Z_{1} \cdot \partial_{Z_{2}} Z_{6} \cdot \partial_{Z_{1}} 
\frac{1}{(2346)} \tilde{\Omega}^{(2)} 
\; = \; \frac{(3461)}{(1234)(2346)}  \,   \tilde{\Omega}^{(1)} \,,
\end{align}
where the (rescaled) one-loop hexagon integral with mixed numerator
is defined as
\begin{align}
\tilde{\Omega}^{(1)}(y_u , y_v , y_w )  \;=& \;  
\frac{(4612)(2346)}{(3461)} \; \int \frac{ d^{4}Z_{AB} } {i \pi^2}
 \frac{  (AB13)(AB (345) \cap (561))}{(AB61)(AB12)(AB23)(AB34)(AB45)(AB56)}
\,.
\end{align}
It is given explicitly by~\cite{ArkaniHamed:2010gh}
\begin{align}
\tilde{\Omega}^{(1)}(y_u , y_v, y_w )  \;=& \; \log u \log v 
- \frac{y_v(1-y_u)}{1-y_u y_v}  \log v \log w 
- \frac{1-y_v}{1-y_u y_v}  \log u  \log w \,.
\end{align}
We note that the integrals $\tilde{\Omega}^{(2)}$ and
$\tilde{\Omega}^{(1)}$ are left invariant by the transformation
\begin{align}
Z_{1} \longleftrightarrow Z_{3} \,,\qquad Z_{4} \longleftrightarrow Z_{6} \,,
\end{align}
which implies 
\begin{align}\label{invariance}
u \longleftrightarrow  v\,,\qquad y_{u} \longrightarrow 1/y_{v}
\,,\qquad y_{v} \longrightarrow 1/y_{u} \,,
\qquad y_{w} \longrightarrow 1/y_{w} \,.
\end{align}
We make the ansatz that $\tilde{\Omega}^{(2)}$ is a pure function, whose
symbol's entries are drawn from the set of nine
letters $\{u,v,w,1-u,1-v,1-w,y_u , y_v ,y_w \}$.  Within this ansatz,
we find that \eqn{diffeqomega2tilde} has a {\it unique}
solution obeying the symmetry condition~(\ref{invariance}),
integrability, and the first entry condition.  The solution involves
parity-even as well as parity-odd terms.

Having determined the symbol of $\tilde{\Omega}^{(2)}$ from the
differential equation~(\ref{diffeqomega2tilde}), we now promote it to
a function. We find that we can express it as\footnote{%
To avoid confusion, we emphasize that $\tilde{V}(y_{v}, y_{w}, y_{u})$
differs from $\mathbb{P} \, \tilde{V}(y_u , y_v , y_w )$, where
$\mathbb{P}$ denotes a cyclic shift of all twistors by one unit.
In fact, we have
$\mathbb{P} \, \{ u, v,w, y_u, y_v, y_w \}
= \{ v, w,u, 1/ y_v, 1/ y_w, 1/ y_u \}$,
and hence 
$\mathbb{P} \, \tilde{V}(y_u , y_v , y_w )
 = - \tilde{V}(y_{v}, y_{w}, y_{u})$. }
\begin{align}\label{result-omega2tilde}
\tilde{\Omega}^{(2)}(y_{u},y_{v},y_{w})
 = \frac{1}{2} \left[ \Omega^{(2)}(v,w,u) + \Omega^{(2)}(w,u,v) \right]
 + Y(u,v,w) + 2 \, \tilde{V}(y_{v},y_{w},y_{u}) \,, 
\end{align}
with 
\begin{align}
Y(u,v,w) = Y^{A}(u,v,w) + Y^{A}(v,u,w) - Y^{B}(u,v,w) \,,
\end{align}
where
\bea
Y^{A}(u,v,w) &=& \frac{1}{2} \biggl\{ 
     4 \, \polylog_4(u) - \polylog_4\left(1-\frac{1}{u}\right)
   + \log u \, \left[ 2 \, \polylog_3(1-u)
                    + 3 \, \polylog_3\left(1-\frac{1}{u}\right) \right]
\nonumber\\ &&\hskip0.3cm\null
   + \frac{2}{3} \, \log^3 u \, \ln(1-u) 
   - \frac{1}{2} \, \left[\polylog_2\left(1-\frac{1}{u}\right) \right]^2
   + \frac{1}{2} \, \log^2 u \, \polylog_2\left(1-\frac{1}{u}\right)
\nonumber\\ &&\hskip0.3cm\null
   - \frac{1}{6} \, \log^4 u
   - 2 \, r(w) + 3 \, \polylog_4\left(1-\frac{1}{w}\right)
\nonumber\\ &&\hskip0.3cm\null
   - \ln\left(\frac{v}{w}\right) \, \left[ 2 \, \polylog_3(1-u)
          + \polylog_3\left(1-\frac{1}{u}\right)
          - \log u \, \polylog_2(1-u) - \frac{1}{6} \, \log^3 u \right]
\nonumber\\ &&\hskip0.3cm\null
   + \frac{1}{2} \, \ln^2\left(\frac{v}{w}\right)
             \, \polylog_2\left(1-\frac{1}{u}\right) \biggr\}
\,,
\eea
with $r(w)$ defined in \eqn{rudef},
and where the beyond-the-symbol ambiguity for a function symmetric
in $u$ and $v$ is given by
\begin{align} \label{beyond-symbol-omegatilde}
Y^{B}(u,v,w) =&\zeta_2  \, \big[  
c_1 (\text{Li}_2(1-u)+\text{Li}_2(1-v))+ c_2 \, \text{Li}_2(1-w)
+ c_3 \left(\log^2 u+\log^2 v \right) \nonumber \\
&\hskip0.5cm
+ \, c_4  \log^2 w + c_5 \log u \log v +  c_6 \log(uv) \log w  \big]
 \nonumber \\
&+  \zeta_3 \, \big[ c_7  \log(uv) + c_8  \log w  \big]  + c_9 \, \zeta_4 \,.
\end{align}
We can ask how many of the $c_{i}$ can be determined by the
differential equation (\ref{diffeqomega2tilde}).  Using the variables
from appendix~\ref{sect-variables} it is not hard to verify that the
only functions appearing in $Y^{B}(u,v,w)$ that are annihilated by the
differential operator are $\zeta_{4}$ and $\zeta_{3} \log (w/(uv)) $.
Therefore, 7 out of the 9 coefficients $c_{i}$ can be determined by
plugging \eqn{result-omega2tilde} back into \eqn{diffeqomega2tilde}.

Indeed, using the parametric integrals derived in the main text for
$\Omega^{(2)}$ and $\tilde{V}$, we can easily verify the differential
equation~(\ref{diffeqomega2tilde}) numerically.  We find
\begin{align}\label{beyond-sol1}
c_{1} = 1 \,, \; c_{2} = -2 \,,\;  c_{3} = {3}/{2} \,,\; 
c_{4} = -1 \,, \;c_{5} = 0 \,, \;c_{6} = 0\,, \; c_{7} = 2- c_{8} \,.
\end{align}
We will fix the remaining two free parameters $c_{8}$ and $c_{9}$ from
boundary conditions that we discuss presently.

\subsection*{Boundary conditions for $\tilde{\Omega}^{(2)}$}

\begin{figure}[t]
\centerline{\includegraphics[scale=0.8]{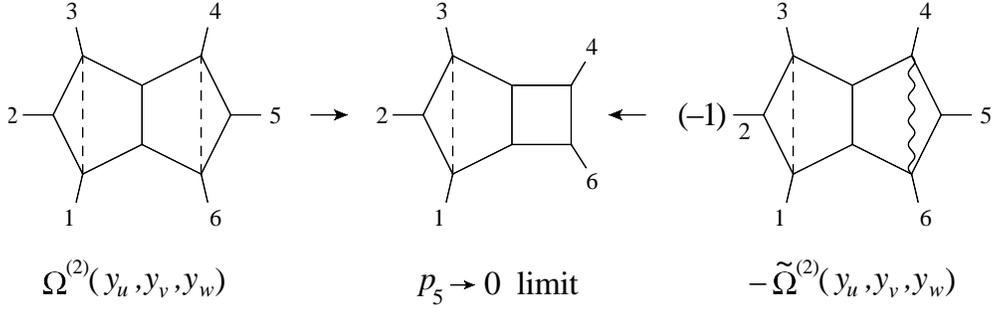}}
\caption{
The integrals ${\Omega}^{(2)}$ and $-\tilde{\Omega}^{(2)}$ have the 
same soft limit $p_{5} \to 0$ at the integrand level. This property
allows us to formulate the boundary condition~(\ref{boundary-condition}).}
\label{fig-soft-limit1}
\end{figure}

Let us discuss appropriate boundary conditions for
$\tilde{\Omega}^{(2)}$.  Here we can use our previous experience with
the integral $\Omega^{(2)}$, which at the integrand level differs from
$\tilde{\Omega}^{(2)}$ only by the numerator in one of the pentagon
subintegrals.  In fact, the numerators of the two integrals are given
by
\begin{align}
N(\tilde{\Omega}^{(2)}) \;=&\; (4612)(2346) (AB13)(CD (561) \cap (345)) \,, \\
N({\Omega}^{(2)}) \;=&\;  (2345)(5612)(3461) (AB13)(CD46) \,.
\end{align}
Previously it was observed that the integrands of these two integrals
reduce to the integrand of a penta-box integral in the soft limit
$p_{5} \to 0$, or equivalently $Z_{5} \to \alpha Z_{4} + \beta Z_{6}$,
as shown in \fig{fig-soft-limit1}~\cite{Drummond:2010mb}.
Unfortunately, the penta-box integral is infrared divergent, so that
the limit is more subtle at the level of integrals.  However, we can use
the fact that the numerator $N(\tilde{\Omega}^{(2)}) +
N({\Omega}^{(2)})$ vanishes linearly in the soft limit. Because the
explicitly known penta-box integral~\cite{Drummond:2010mb}
only has logarithmic divergences, we expect the
following boundary condition to hold,
\begin{align}\label{boundary-condition}
\lim_{\tau \to 0} \;( \tilde{\Omega}^{(2)} 
+ {\Omega}^{(2)})(\xi_1 \tau, \xi_2  \tau, 1- \tau ) = 0\,.
\end{align}
Here we have parametrized the soft limit for the cross-ratios $u,v,w$
by $\tau \to 0$.\footnote{There is a slight abuse of notation here
since, strictly speaking, $\tilde{\Omega}^{(2)}$ should be thought of
as a function of the $y$ variables. However, in the soft limit,
its parity-odd piece vanishes, justifying the use of the $u$ variables.}
We have verified equation (\ref{boundary-condition}) at the symbol
level. In the following we will assume it holds also at the level of
functions.

A related observation is that $\Omega^{(2)}$ vanishes in cyclically
related soft limits,
\begin{align}
\lim_{\tau \to 0} \; \Omega^{(2)}(1- \tau , \xi_1 \tau, \xi_2  \tau ) = 0 \,.
\end{align}
This vanishing can in fact be understood as a property of the pentagon
sub-integral.  Since $\tilde\Omega^{(2)}$ contains the same
sub-integral as $\Omega^{(2)}$, we expect the same boundary condition
to hold, {\it i.e.}
\begin{align}\label{rotated-soft-limit}
\lim_{\tau \to 0} \; \tilde\Omega^{(2)}(1- \tau , \xi_1 \tau, \xi_2  \tau )
  = 0\,.
\end{align}

We find that imposing the two boundary
conditions~(\ref{boundary-condition}) and (\ref{rotated-soft-limit})
fixes all but one of the beyond-the-symbol ambiguities in
\eqn{beyond-symbol-omegatilde},
\begin{align}\label{beyond-sol2}
c_1 = 1-{c_9}/{5}\,, \; c_2 =  -{c_9}/{5}-2 \,, \; c_3 =  {3}/{2}\,,\;
c_4 = -1 \,, \;c_5 =0 \,,\;  c_6 = 0 \,, \; c_7 = 2 \,,  \; c_8 = 0\,.
\end{align}
Comparing to \eqn{beyond-sol1}, we see that the two solutions are
compatible with each other, which is a non-trivial cross check.
Moreover, taken together they uniquely fix all the beyond-the-symbol
parameters, and we have finally,
\begin{align}\label{beyond-sol3}
c_{1} = 1 \,, \; c_{2} = -2 \,, \; c_{3} = {3}/{2} \,, \; 
c_{4} = -1 \,,\;  c_{5} = 0 \,,\;  c_{6} = 0\,, \; 
c_{7} = 2 \,,  \; c_{8} = 0\,,  \; c_{9}=0   \,.
\end{align}


\section{Useful variables}
\label{sect-variables}
\setcounter {equation} {0}

In this paper, we found it useful to work with several sets of variables.
We can express the letters appearing in our symbols in terms of
four-brackets of twistors,
\begin{align}
u = & \, \frac{(6123)(3456)}{(6134)(2356)} \,,\qquad  
v = \frac{(1234)(4561)}{(1245)(3461)} \,,\qquad 
w = \frac{(2345)(5612)}{(2356)(4512)} \,,
\\
1-u = & \, \frac{(1356)(2346)}{(1346)(2356)} \,, \quad
1-v = \frac{(2461)(3451)}{(2451)(3461)} \,, \quad
1-w = \frac{(3512)(4562)}{(3562)(4512)} \,, 
 \\
y_{u} = & \, \frac{(2361)(2456)(3451)}{(2351)(2461)(3456)} \,,\ \ 
y_{v} =  \frac{(3462)(3512)(4561)}{(3412)(3561)(4562)} \,,\ \ 
y_{w} =  \frac{(1246)(1356)(2345)}{(1256)(1345)(2346)} \,.
\end{align}
Since the twistors are redundant, it can sometimes be useful to have
a particular para\-metri\-zation for them, {\it e.g.}
\begin{align}
Z_{1} =& (1,1,\gamma,1) \,, \qquad Z_{2}=(1,0,0,0) 
\,, \qquad Z_{3} = (0,1,0,0) \,, \nonumber 
\\ Z_{4} =& (0,0,1,0) \,, \qquad Z_{5} = (0,0,0,1) 
\,, \qquad Z_{6} = (1,\alpha,1,\beta) \,,
\end{align}
with 
\begin{align}
\alpha = \frac{1 - y_u y_v y_w}{1 - y_v y_w} \,,
\quad \beta = \frac{1 - y_u y_v y_w}{1 - y_w} \,,
\quad \gamma = \frac{1 - y_w}{1 - y_u y_w} \,.
\end{align}
Although the $y$ variables are constructed using square roots
of the original cross ratios $u$, $v$ and $w$, the cross ratios
themselves are rational combinations of the variables $y_u$, $y_v$
and $y_w$.  The explicit relations are,
\begin{align}
& u = \frac{y_u(1-y_v)(1-y_w)}{(1-y_w y_u)(1-y_u y_v)} \,, \quad
v = \frac{y_v (1-y_w) (1-y_u)}{(1-y_u y_v) (1-y_v y_w)} \,, \quad
w = \frac{y_w (1-y_u) (1-y_v)}{(1-y_v y_w) (1-y_w y_u)} \,,~~ \\
& 1-u = \frac{(1-y_u) (1-y_u y_v y_w)}{(1-y_w y_u) (1-y_u y_v)} \,, \qquad
   1-v = \frac{(1-y_v) (1-y_u y_v y_w)}{(1-y_u y_v) (1-y_v y_w)} \,, \\
& 1-w = \frac{(1-y_w) (1-y_u y_v y_w)}{(1-y_v y_w) (1-y_w y_u)} \,, \qquad
\sqrt{\Delta} = \frac{(1-y_u) (1-y_v) (1-y_w) (1-y_u y_v y_w)}
                       {(1-y_u y_v) (1-y_v y_w) (1-y_w y_u)} \,,
\label{rtdeltayrelation}
\end{align}
where we have picked a particular branch of $\sqrt{\Delta}$.


\bibliographystyle{nb}
\bibliography{nmhv}

\end{document}